\documentclass[11 pt]{article}
\usepackage[latin1]{inputenc}
\usepackage{graphicx}
\usepackage{float}
\usepackage{caption}
\usepackage{subcaption}
\graphicspath{{figures/}}

\usepackage[title]{appendix}

\usepackage{hyperref}
\usepackage{nicefrac}

\usepackage[mathscr]{eucal}
\usepackage{amsmath}
\usepackage{amssymb}
\usepackage{musicography}
\usepackage{braket}
\usepackage{cancel}
\usepackage[colorinlistoftodos]{todonotes}

\usepackage{color}

\newcommand{\fref}[1]{Fig. \ref{#1}}
\newcommand{\sref}[1]{Sec. \ref{#1}}

\usepackage{xcolor}
\definecolor{cblue}{RGB}{81, 167, 192}
\definecolor{cpurple}{RGB}{136, 60, 78}

\title{
{Approximate Killing symmetries in non-perturbative quantum gravity}
}

\begin{document}
\vspace{20pt}
\begin{center}
{\Large\bf Approximate Killing symmetries in\\ non-perturbative quantum gravity\\
}
\vspace{15pt}
{\large J.\ Brunekreef$^{a, \sharp}$, M.\ Reitz$^{a, \flat}$}
\vspace{15pt}

$^{a}${\sl Institute for Mathematics, Astrophysics, and Particle Physics, Radboud University, }\\
{\sl Heyendaalseweg 135, 6525 AJ Nijmegen, The Netherlands }
\vspace{15pt}

emails:  {\sl 
$^{\sharp}$jorenb@gmail.com,
$^{\flat}$m.reitz@science.ru.nl}

\vspace{40pt}

\begin{abstract}
It is an open question whether fluctuations at the Planck scale in a non-perturbative theory of quantum gravity behave in such a way that the resulting semi-classical geometry can be modelled by a space that admits (approximate) Killing symmetries. We have investigated whether the notion of approximate Killing vector fields is suitable to address this question in lattice theories of quantum gravity, such as (Causal) Dynamical Triangulations. We show that it is possible to construct quantum observables related to approximate Killing vector fields using the framework of Discrete Exterior Calculus. We have evaluated the expectation value of one particular choice of observable on three toy models of two-dimensional quantum gravity.
\end{abstract}
\today
\end{center}

\newpage
\tableofcontents
\newpage
\section{Introduction}
\label{sec:motivation}

A central question in theories of quantum gravity is whether an emerging semi-classical geometry, if present at all, can be well approximated by a space that is close to homogeneous and isotropic, even though the model may contain large fluctuations at the Planck scale. (Causal) Dynamical Triangulations is a non-perturbative approach to quantum gravity, based on a lattice regularisation of space-time, in which these kinds of questions can possibly be addressed. Homogeneous and isotropic space-times can (in part) be characterised by the existence of Killing vector fields. In the real universe, these symmetries only exist approximately. Formulating methods to describe the presence of approximate Killing symmetries in models of dynamical geometry could be an interesting route to understanding gravity on different scales. We expect that a notion of approximate symmetry should emerge from a physical quantum theory of gravity at some large scale in the theory. In this article we will investigate the suitability of a specific definition of approximate Killing vectors for the construction of observables in non-perturbative theories of quantum gravity based on lattice regularisations of space-time. The definition of approximate Killing vector fields we consider can be generalised to simplicial manifolds. In the context of Regge calculus, this generalisation may be relevant for both the classical averaging problem and for various models of quantum gravity.

The precise notion of approximate symmetry we will use is the $\lambda$-approximate Killing vector, a concept introduced by Matzner \cite{Matzner:1967}. The notion of approximate symmetries has until now only been considered in a classical context. In the work of Matzner, $\lambda$-approximate Killing vectors were used to calculate the energy content of gravitational radiation with small amplitudes. More recently, a related definition of approximate symmetries was used to define conserved Komar currents related to the energy of gravitational radiation \cite{Feng:2018}. Various definitions of approximate Killing vectors have also been used to calculate the angular momentum of black hole binary systems \cite{Cook:2007}. This work is different because we are not primarily interested in the approximate symmetries of a single geometry, but in approximate symmetries of an ensemble average of geometries in a non-perturbative theory of quantum gravity. To the best of our knowledge, $\lambda$-approximate symmetries have not been studied in the context of quantum gravity before.

The models  in which we have studied the possible presence of $\lambda$-approximate Killing vector fields are Causal Dynamical Triangulations (CDT), Dynamical Triangulations (DT), and a model adapted from CDT which we call small perturbations around flat space. We will describe the small perturbations around flat space in Sec.\ \ref{ssec:results-discussion}. We will only discuss the elements of CDT and DT that are necessary to understand the context of our investigations of approximate symmetries in quantum gravity. For a more thorough exposition we refer the reader to the review articles on these approaches \cite{Loll:2012, Loll:1998, Loll:2019}.

In short, CDT and DT are non-perturbative formulations of quantum gravity, defined in terms of a lattice regularisation of respectively a Lorentzian and Euclidean gravitational path integral. DT and, after a Wick rotation, CDT take the form of a sum over Euclidean piecewise flat geometries or ``triangulations" built from simplices (the generalisation of triangles to arbitrary dimension) with a characteristic length scale $a$. The length scale $a$ serves as a UV cut-off of the path-integral and a finite number $N$ of simplices in a triangulation $T$ takes the role of an IR regulator. These piecewise flat geometries are called simplicial manifolds and they form a particular class of simplical complexes. In Sec.\ \ref{sec:dec}, we give a more detailed definition of these geometries.

CDT and DT differ in terms of the configuration space of geometries that appear in their respective state sums. The configuration space $\mathcal{T}_{DT}$ of DT is, for a particular choice of $n$-dimensional manifold $\mathcal{M}$, given by all possible gluings of equilateral $n$-dimensional simplices that are homeomorphic to $\mathcal{M}$. All link lengths in DT geometries are equal to $a$. The configuration space $\mathcal{T}_{CDT}$ of CDT is defined similarly, but comes with a layered structure that is inherited from the causal structure of the Lorentzian metric. CDT geometries consist of multiple layers of $(n-1)$-dimensional DT geometries at constant integer time that are connected by links. This is done in such a way that the layers are connected by $n$-dimensional simplices and the topology of the geometry does not change along this layered structure. The length of the links within the layers can differ from the length of the links between different layers. However, in this text we will only consider the case where all links are of equal length $a$.

These lattice regularisations allow for the application of numerical methods to study the properties of the quantum theory. The physical properties of CDT and DT can be studied through Monte Carlo methods. With a Monte Carlo algorithm weighted with the exponentiated Regge action $e^{-S_R[T]}$, the analogue of the Einstein-Hilbert action for a simplicial manifold $T$ \cite{Regge:1961}, a quantum ensemble of geometries can be sampled. The finite resources available for such numerical methods make it necessary to conduct these simulations at a finite number of simplices $N$. Often the geometries are chosen to be closed simplicial manifolds. The infinite volume limit can then be approximated by repeating these simulations at increasing values of $N$ and extrapolating to $N\rightarrow \infty$.

Physical properties can be studied by calculating expectation values of suitable observables on the sampled ensemble. The expectation values can be approximated by an average over the generated geometries if an infinite volume limit can be extrapolated from the scaling of the observables. When studying observables, some care needs to be taken with respect to the lattice cut-off $a$ and the finite system size $N$. Both are non-physical regulators, and properties of observables that are sensitive to length scales related to either $a$ or $N$ should not be interpreted as physical. Such properties are called lattice artefacts and finite-size effects, respectively.

In CDT, several observables have been formulated for which the expectation value in a classical limit is in agreement with general relativity. Furthermore, some of these observables give non-trivial predictions near the Planck scale. These observables include the spectral and Hausdorff dimensions, the volume profile, and the quantum Ricci curvature \cite{Loll:2019}. However, a better understanding of the geometric properties of CDT requires observables that give more detailed geometric information. Of particular interest are observables that contain directional information. Recent investigations of the quantum Ricci curvature show promising signs that directional properties can be studied \cite{qrc1,qrc2,qrc3} and it would be beneficial to find independent observables to support these studies. We also need more observables that can determine whether any sensible (semi-)classical behaviour is present in the theory and can probe different scales. An observable that can tell us about the presence of large-scale symmetries can be very valuable with regard to these questions. In this article we develop tools to study approximate symmetries in models of quantum gravity based on a lattice regularisation of space-time. We have investigated the construction of potential observables from $\lambda$-approximate Killing vectors.

To study the notion of approximate symmetries in the context of DT and CDT we need to generalise the definition of $\lambda$-approximate Killing vector fields to a discrete setting. This can be achieved by using the framework of Discrete Exterior Calculus (DEC). DEC is a formulation of exterior calculus on simplicial complexes. The goal is to establish whether we can define an observable which can be used to investigate whether at larger scales an effective space-time with a certain number of approximate symmetries emerges from the quantum ensembles in these models of quantum gravity. As a proof of concept, we propose a particular observable related to approximate Killing vector fields. Furthermore, we calculate the expectation value of this observable in three simple two-dimensional models of random geometry.

We will argue that the discrete approximate Killing vectors show promise as an ingredient to construct observables to study effective symmetries in quantum gravity when fluctuations are small. The main result of this article is a comparison between the three different two-dimensional toy models of quantum geometry, with respect to the proposed observable. In \sref{ssec:killing-energy}, we describe the exact definition of $\lambda$-approximate Killing vector fields. \sref{app:bochner} discusses a reformulation of the $\lambda$-approximate Killing vector fields in terms of an eigenvalue problem for the case of two dimensions. Secs. \ref{sec:dec} and \ref{sec:dkvf} summarise the necessary ingredients from the framework of Discrete Exterior Calculus that are needed to generalise $\lambda$-approximate Killing vector fields to simplicial manifolds. In \sref{sec:example-geometries}, we discuss properties of discrete approximate Killing vector fields (DAKVFs) on the discrete analogues of manifolds that admit exact Killing vector fields. The proposed observable is introduced in \sref{sec:symmetries-qg}. Finally, we present and discuss the results of a measurement of the proposed observable on two-dimensional toy models of quantum gravity based on DT, CDT and small perturbations on flat space of toroidal topology.

\section{\texorpdfstring{$\lambda$}{lambda}-approximate Killing vector fields}
\subsection{Killing energy}
\label{ssec:killing-energy}
In order to investigate whether some notion of approximate symmetry is present in the previously mentioned discrete models of quantum gravity, we first need to define a notion of approximate symmetries. To set the stage, first note that by symmetries we mean the isometries generated by the Killing vector fields of an $n$-dimensional differentiable manifold $\mathcal{M}$, endowed with a smooth metric $g_{\mu \nu}$ of Euclidean signature. In this article, we will only consider closed manifolds $\mathcal{M}$. We consider metrics of Euclidean signature because they are the type of geometries that are relevant for CDT, as was described in \sref{sec:motivation}. A one-form $\xi$ dual to a Killing vector field is defined by the Killing equation, which is equivalent to a vanishing Lie derivative $\mathcal{L}_\xi$ of the metric along $\xi$,
\begin{equation}
\mathcal{L}_\xi g_{\mu\nu}=\nabla_\mu \xi_\nu + \nabla_\nu \xi_\mu =0.
\label{eq:killing-equation}
\end{equation}
From now on, we will often use a Killing vector field $\xi^\mu$ and its dual one-form $\xi_\mu$ interchangeably. Furthermore, we will understand `Killing vector fields' and `Killing vectors' to be one and the same.

There exist a few different definitions of a generalisation of $\xi$ to approximate Killing vectors. A comparison of various definitions of approximate Killing vectors can be found in \cite{Approximate symmetries in General relativity}. We will choose a definition based on what is called the Killing energy $E(\omega)$ of a general one-form $\omega$. The original formulation of this definition of approximate Killing vectors goes back to Matzner \cite{Matzner:1967}. For a general one-form $\omega$ we introduce the notation $K_{\mu \nu}$ for the Lie derivative $\mathcal{L}_\omega$ of the metric:
\begin{equation}
K_{\mu\nu}\equiv\mathcal{L}_\omega g_{\mu\nu} =\nabla_\mu \omega_\nu + \nabla_\nu \omega_\mu.
\label{eq:lie-d}
\end{equation}
The contraction of $K_{\mu\nu}$ with itself is positive semi-definite for Euclidean signature,
\begin{equation}
K_{\mu\nu}K^{\mu\nu}\geq 0.
\label{eq:tensor-norm}
\end{equation}

This lower bound is attained if and only if $\omega$ is dual to a Killing vector field such that the Killing equation \eqref{eq:killing-equation} is satisfied. From eq.\ \eqref{eq:tensor-norm} we define the Killing energy $E(\omega)$ of a one-form $\omega$ by an integral over the $n$-dimensional manifold $\mathcal{M}$. We will denote the value of the Killing energy by $\lambda$, so that
\begin{equation}
E(\omega)=\int_\mathcal{M} dV \ K_{\mu\nu}K^{\mu\nu}  = \lambda,
\label{eq:killing-energy}
\end{equation}
where $dV=\sqrt{g}~d^nx$ is the volume element of $\mathcal{M}$. The value $\lambda$ is also positive semi-definite and is zero if and only if $\omega$ is a one-form dual to a Killing vector field. Finding a minimum of the Killing energy $E(\omega)$ with respect to $\omega$ with value zero is therefore equivalent to solving the Killing equation. In general, $\mathcal{M}$ does not admit any exact Killing vector fields. The Killing energy $E(\omega)$ can however also be minimised when no such vector field exists on $\mathcal{M}$. If the minimal value of $\lambda$ for a manifold $\mathcal{M}$ is sufficiently small, we can expect the corresponding vector field $\omega$ to be close to an exact Killing vector field of a manifold $\mathcal{M'}$ which does admit an exact Killing symmetry and can be obtained by a small deformation of $\mathcal{M}$ \cite{Beetle:2013}. Following \cite{Ben-Chen:2010} we will call the smooth vector field dual to a one-form that minimises the Killing energy a $\lambda$-approximate Killing vector. A $\lambda$-approximate Killing vector can be seen as the generator of an ``almost'' isometry of $\mathcal{M}$, in the sense that the variation of the metric along the flow of the $\lambda$-approximate Killing vector is small if $\lambda \ll \sqrt{R_{\kappa\lambda\mu\nu}R^{\kappa\lambda\mu\nu}}$ \cite{Matzner:1967}. This bound on $\lambda$ is related to the size of the patch for which Riemann normal coordinates are valid \cite{Nesterov:1999}. We will not give a precise definition of what is meant by ``almost'' isometry here. If we only consider normalised one-forms $\omega$, i.e.
\begin{equation}
\int_\mathcal{M} dV \omega_\mu\omega^\mu  = 1,
\end{equation}
$\lambda$-approximate Killing vectors are uniquely defined. The Killing energy can be rewritten, by use of a variation of the Bochner technique, in terms of the exterior derivative $(d\omega)_{\mu \nu}=2\nabla_{[\mu}\omega_{\nu]}$, the co-differential $\delta\omega=\nabla^\mu \omega_\mu$ of $\omega$ and the Ricci tensor $R_{\mu \nu}$ contracted with $\omega$,
\begin{equation}
E(\omega) = \int_\mathcal{M}dV \ \left( 2|d \omega,d \omega| +4 |\delta \omega,\delta\omega| - 2R_{\mu\nu} \omega^\mu\omega^\nu\right),
\label{eq:killing-energy-hodged}
\end{equation}
where we use the inner product for scalars $|\phi,\phi|=\phi^2$, one-forms $|\omega,\omega|=\omega_{\mu}\omega^{\mu}$ and two-forms $|\psi,\psi|=\frac{1}{2}\psi_{\mu\nu}\psi^{\mu\nu}$. The operators $d$ and $\delta$ are adjoint to each other with respect to the inner product $|\cdot,\cdot|$, which fixes the normalisation by $\frac{1}{2}$ in the inner product on two-forms. For general manifolds $\mathcal{M}$, this expression of $E(\omega)$ would include a boundary term, which was omitted here because we will only consider closed manifolds $\mathcal{M}$ (see appendix \ref{app:bochner} for details of the derivation).

Up to this point the discussion was valid for arbitrary dimension $n$. In this work, we will only consider the case of two-dimensional closed manifolds, as this is sufficient for the toy models of quantum gravity that we discuss. Work on $\lambda$-approximate Killing vectors in higher dimensions is under way.

\subsection{The Killing vector in two dimensions as an eigenvalue problem}
\label{ssec:killing-eigen-problem}

In two dimensions the Ricci tensor $R_{\mu\nu}$ reduces to $R_{\mu\nu}=\frac{R}{2}g_{\mu \nu}$, in terms of the Ricci scalar $R$. The Killing energy simplifies to
\begin{equation}
E(\omega) = \int_\mathcal{M} dV \ \left(2| d \omega,d \omega| + 4 | \delta \omega,\delta \omega | - 2 R | \omega,\omega |\right).
\label{eq:killing-energy-hodge}
\end{equation}
We can now write $E(\omega)$ in a more compact form,
\begin{equation}
E(\omega) = \int_\mathcal{M} dV \ |S \omega, \omega|,
\label{eq:killing-energy-s}
\end{equation}
with
\begin{equation}
S = 2\delta d + 4 d \delta - 2R.
\label{eq:killing-operator}
\end{equation}
From eq.\ \eqref{eq:killing-energy} we see that the Killing energy $E(\omega)$ is equal to $\lambda$ for a one-form $\omega$ when
\begin{equation}
S\omega=\lambda\omega.
\label{eq:eigenvalue-problem}
\end{equation}
Solving eq.\ \eqref{eq:eigenvalue-problem} is now an eigenvalue problem for $S$. Because $S$ is positive semi-definite, minimising the Killing energy with respect to $\omega$ is equivalent to finding the smallest eigenvalue $\lambda_0$ that solves the eigenvalue problem \eqref{eq:eigenvalue-problem}. The eigenvector $\omega_0$ corresponding to $\lambda_0$ is the one-form that minimises the Killing energy and is dual to a $\lambda$-approximate Killing vector. The lowest eigenvalue $\lambda_0$ is equal to zero for the special case that $\mathcal{M}$ admits an exact Killing vector field.

A $n$-dimensional manifold has $\frac{n(n+1)}{2}$ independent Killing vector fields if it is maximally symmetric. In our specific case of two dimensions, the operator $S$ therefore has three degenerate eigenvectors if $\mathcal{M}$ is maximally symmetric. The global topology can impose additional restrictions on the number of Killing vectors. For example, a manifold with the topology of a two-torus can have at most two Killing vectors. We can therefore expect situations where it is sensible to define multiple $\lambda$-approximate Killing vectors, depending on the topology and dimension of the manifold under consideration. The interpretation of a $\lambda$-approximate Killing vector in terms of a generator of ``almost" isometries is however only well-understood when the corresponding eigenvalue of $S$ is close to zero. Our ultimate goal, as described in \sref{sec:motivation}, is to study the notion of $\lambda$-approximate Killing vectors in non-perturbative quantum gravity. The models that we will investigate are based on a lattice regularisation in which the underlying manifolds are piecewise linear simplicial complexes.

The next step is to find a discrete counterpart of $S$ defined in eq.\ \eqref{eq:killing-operator}, which is applicable in a piecewise flat context. In work by Ben-Chen et al. \cite{Ben-Chen:2010} it was shown that this can be achieved in the framework of Discrete Exterior Calculus (DEC). We will use their definition of the discrete analogue of $S$. The following section contains a summary of the necessary elements relevant to this article.

\section{Exterior calculus on simplicial manifolds}
\label{sec:dec}

\begin{figure}
\centering
\includegraphics[width=0.60\linewidth]{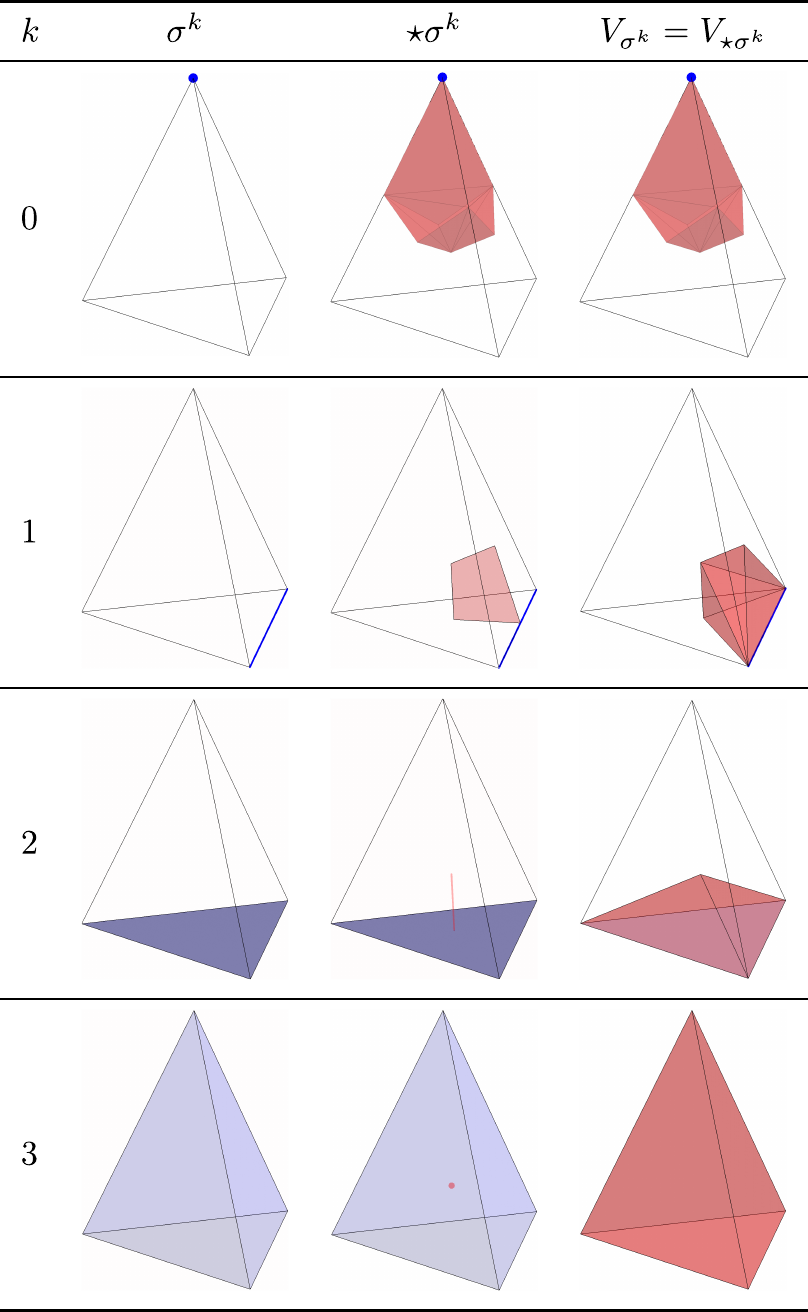}
\caption{The various $k$-dimensional primal sub-simplices $\sigma^k$, dual sub-simplices $\star\sigma^k$ and support volumes $V_{\sigma^k}$ and $V_{\star\sigma^k}$ are shown for $n=3$. The primal simplices are shown in blue and the corresponding dual simplices and support volumes in red.}
\label{fig:3-simplices}
\end{figure}

In this section we will present a summary of the framework that we will use for defining exterior calculus on the piecewise linear simplicial complexes of DT and CDT. The framework is called Discrete Exterior Calculus (DEC). A thorough review can be found in \cite{Discrete Exterior Calculus}. An $n$-dimensional piecewise flat oriented simplicial complex $T$ is given by a collection of oriented simplices $\sigma^k$, with $k \in \{ 0, ... , n \}$. We write $N_k$ for the number of $k$-simplices in $T$. The interior of the simplices is endowed with a Euclidean flat metric. A simplicial complex is defined such that every subsimplex of $\sigma^n \in T$ is also part of $T$. Also, the intersection of two simplices $\sigma^k,\bar{\sigma}^{k'}$ is a subsimplex of both $\sigma^k$ and $\bar{\sigma}^{k'}$. We furthermore restrict our discussion to simplicial complexes for which every $\sigma^k$ with $k<n$ is contained in some $\sigma^n\in T$, and we write $\sigma^k \prec\sigma^n$. Every simplex $\sigma^k$ is the convex hull $[\sigma_0^0,...,\sigma^0_{k}]$ of $k+1$ vertices $\sigma^0$ and contains $\binom{k+1}{k'+1}$ simplices $\sigma^{k'}$, $k'<k$. The orientation is defined by the labelling $v_i$ of the vertices $\sigma_{v_i}^0$ contained in the simplex $\sigma^k$. We choose the simplex $\sigma^k=[\sigma_{v_0}^0,...,\sigma^0_{v_i},...,\sigma^0_{v_k}]$ to be positively oriented for $v_0<v_i<v_k$. The relative sign of a simplex $\sigma^k$ and a sub-simplex $\sigma^{k'}$, with $k>k'$, is given by $\textrm{sign}(\sigma^k ;\sigma^{k'}) =\textrm{sgn}(\pi)$, where $\textrm{sgn}(\pi)$ is the sign of the permutation $\pi$ of the labels with which the subsimplex $\sigma^{k'}=[\sigma_{u_0}^0,...,\sigma^0_{u_i},...,\sigma^0_{u_k'}]$ can be embedded in $\sigma^k$,
\begin{equation}
[\sigma_{\pi(v_0)}^0,...,\sigma^0_{\pi(v_i)},...,\sigma^0_{\pi(v_{k})}]=[\sigma_{v_0}^0,...,\sigma^0_{v_{k-k'}},\sigma_{u_0}^0,...,\sigma_{u_{k'}}^0].
\label{eq:simplex-orientation}
\end{equation}
The simplicial complexes that we consider are also manifolds in the sense that they are locally homeomorphic to $\mathbb{R}^n$ and are also called piecewise linear manifolds or simplicial manifolds. The numbers $N_k$ are not independent, but are related by the Euler characteristic $\chi(T)$. For an $n$-dimensional simplicial complex $T$, the Euler characteristic is defined as
\begin{equation}
\chi(T)= N_0-N_1+N_2 + ... + (-1)^n N_n.
\label{eq:euler-characteristic}
\end{equation}
We furthermore define $c(\sigma^k)$ to be the center of the circumsphere (the circumcenter) of the $k+1$ vertices contained in $\sigma^k$. The circumcenter of a general simplex does not necessarily lie within that simplex. For simplicity, we will exclude such geometries from our analysis. Simplicial manifolds that only contain simplices that contain their own circumcenter are called circumcentric. The simplicial manifolds relevant to the discrete models of quantum gravity we have studied are circumcentric. Using the circumcenter we define the dual complex $\star T$. The dual complex consists of the collection of $(n-k)$-dimensional polygons\footnote{The polygons $\bar{\sigma}^{n-k}$ are constructed as the formal sum $\sum _{\sigma^k\prec...\prec\sigma^{k'}\prec...\prec\sigma^n}$ of the $k'$-simplices that simultaneously contain $\sigma^k$ and are contained in $\sigma^n$. The sum should be read as the union of the simplices $[c(\sigma^k),...,c(\sigma^n)]$, taking into account the induced orientation.} $\bar{\sigma}^{n-k}=\star\sigma^k$, dual to a $k$-simplex $\sigma^k$,
\begin{equation}
\star\sigma^k=\sum _{\sigma^k\prec...\prec\sigma^n}\epsilon_{\sigma^k,...,\sigma^n}[c(\sigma^k),...,c(\sigma^n)]. 
\label{eq:dual-simplex}
\end{equation}
To differentiate between the two discrete spaces we will call the simplicial complex $T$ the primal complex. The totally antisymmetric symbol $\epsilon_{\sigma^k,...,\sigma^n}$ ensures that the dual complex has the correct orientation induced by the primal complex, 
\begin{equation}
\epsilon_{\sigma^k,...,\sigma^n}=\textrm{sign}[[c(\sigma^0),...,c(\sigma^k)];\sigma^k]\cdot\textrm{sign}[[c(\sigma^0),...,c(\sigma^n)];\sigma^n],
\label{eq:dual-orientation}
\end{equation}
where the relative signs between dual and primal simplices are defined analogously to eq.\ \eqref{eq:simplex-orientation}. For the top-dimensional case (in which $k = n$), eq.\ \eqref{eq:dual-simplex} identifies the circumcenter of a simplex $\sigma^n$ with its dual, i.e. $\star\sigma^n=c(\sigma^n)$. Another useful object is the support volume $V_{\sigma^k}$ of a simplex $\sigma^k$. It is given by the convex hull of $\sigma^k$ and its dual $\star\sigma^k$. These objects are illustrated in \fref{fig:3-simplices} for the case $n=3$ in a tetrahedron.

We now have discussed all the objects that define an $n$-dimensional simplicial complex $T$ and the associated dual complex $\star T$ based on the circumcenters of $T$. For our purposes, we also need a discrete counterpart of the operators of exterior calculus. Construction of such generalisations is notoriously ambiguous \cite{Discrete Exterior Calculus}, with various advantages and disadvantages specific to each approach. One such approach is the framework of Discrete Exterior Calculus (DEC), which turns out to be particularly suitable for constructing discrete approximate Killing vector fields. The main advantage of DEC for our purposes is that it is a discrete implementation of Hodge theory. All the operators that appear in eq.\ \eqref{eq:killing-operator} are straightforward to define in this framework. We will now give a short summary of the definitions of these operators in DEC.

The first ingredient is the discrete counterpart of a differential form. Discrete differential forms are defined as functions on the free Abelian group $C^k(T;\mathbb{Z}_2)$ of formal sums of $k$-simplices, $c^k \in C^k(T;\mathbb{Z}_2)$,
\begin{equation}
c^k=\sum_i a_i c^k_i,
\end{equation}
for which the basis elements $c^k_i$ are the set of $k$-simplices $\sigma^k \in T$ and $a_i$ are taken from the values $\{0,1\}$, i.e. from the cyclic group $\mathbb{Z}_2$. A discrete $k$-form $\alpha^k$ is an element of the space of homomorphisms $\mathrm{Hom}(C^k(T),\mathbb{R})\equiv\Omega^k(T)$ between $C^k(T;\mathbb{Z}_2)$ and $\mathbb{R}$. It is defined through the natural pairing\footnote{This notation was chosen on purpose to be similar to the notation for the inner product in eq.\ \eqref{eq:killing-energy-hodge}. The natural pairing $\langle\cdot,\cdot\rangle$ is the natural discrete analogue of the continuum inner product $|\cdot,\cdot|$.} $\langle\cdot,\cdot\rangle$,
\begin{equation}
\langle \alpha^k,c^k \rangle \equiv \alpha(c) = a\in \mathbb{R},
\label{eq:natural-pairing}
\end{equation}
which should be read as the evaluation of $\alpha^k$ on a formal sum $c^k$ of $k$-simplices $\sigma^k$. Eq.\ \eqref{eq:natural-pairing} can also be expressed in terms of a basis for both $\Omega^k(T)$ and $C(T;\mathbb{Z}_2)$. We will denote the basis elements of $\Omega^k(T)$ by $\alpha^k_{i}$. The natural pairing on the basis elements is given by
\begin{equation}
\langle \alpha^k_{i}, \sigma_j^k \rangle = \delta_{ij}.
\label{eq:basis-pairing}
\end{equation}
Here the subscripts $i$ and $j$ label the $k$-simplices and will be suppressed for expressions for an arbitrary $k$-simplex $\sigma^k$. With the basis $\alpha^k_i$, the $k$-forms $\alpha^k$ can be represented as a vector of dimension equal to the number of links in the triangulation $T$. The connection to the continuum $k$-form $\alpha_c^k$ is made by defining 
\begin{equation}
\langle \alpha^k, \sigma^k \rangle = \int_{\sigma^k}\alpha_c^k,
\label{eq:integration}
\end{equation}
with respect to the flat metric on $\sigma^k$. A boundary operator $\partial$ acting on a $k$-simplex $\sigma^k$ is defined by
\begin{equation}
\partial_k\sigma^k=\sum_{\sigma^0_i\prec\sigma^k}(-1)^i[\sigma^0_0,...,\cancel{\sigma^0_i},...,\sigma^0_k],
\label{eq:boundary}
\end{equation}
where the summation runs over the $k+1$ vertices contained in $\sigma^k$ and $\cancel{\sigma^0_i}$ denotes the omission of vertex $\sigma^0_i$. The boundary operator $\partial_k:C^{k+1}(T;\mathbb{Z}_2) \rightarrow C^k(T;\mathbb{Z}_2)$ is a homomorphism between the groups $C^k(T;\mathbb{Z}_2)$. The boundary operator is nil-potent, i.e. $\partial^k\circ\partial^{k+1}=0$. The definition of the discrete exterior derivative $d_k :\Omega^{k}(T) \rightarrow \Omega^{k+1}(T)$ is induced by the boundary operator $\delta$. The discrete exterior derivative $d_k$ is a map from the space of discrete $k$-forms $\Omega^{k}(T)$ into the space of discrete $(k+1)$-forms $\Omega^{k+1}(T)$ and is given explicitly by the relation
\begin{equation}
\langle d_k \omega^k, \sigma^{k+1} \rangle = \langle \omega^k,\partial_k \sigma^k \rangle.
\label{eq:exterior-derivative}
\end{equation}

In the remainder of the text, we will omit the subscript $k$ on $d_k$ and $\delta_k$ and take it to be implied from the context. The construction on the dual complex is analogous, defining formal sums $c^k \in C^k(\mathcal{\star T};\mathbb{Z}_2)$ of the dual $k$-simplices $\star\sigma^k$ and dual $k$-forms $\Omega^k(\star T)$ that act on the dual $k$-simplices. The boundary operator on the dual simplex $\bar{\sigma}^k$ is given by
\begin{equation}
\partial\bar{\sigma}^k=\partial\star\sigma^{n-k}=\sum_{\sigma^{n-k+1}\succ\sigma^{n-k}}\textrm{sign} (\sigma^{n-k+1};\sigma^{n-k})\star\sigma^{n-k+1}.
\label{eq:dual-boundary}
\end{equation}

\begin{figure}[t]
\centering
     \begin{minipage}[t]{\textwidth}
      \centering
\def\svgwidth{0.6\columnwidth}
\tiny{
\includegraphics{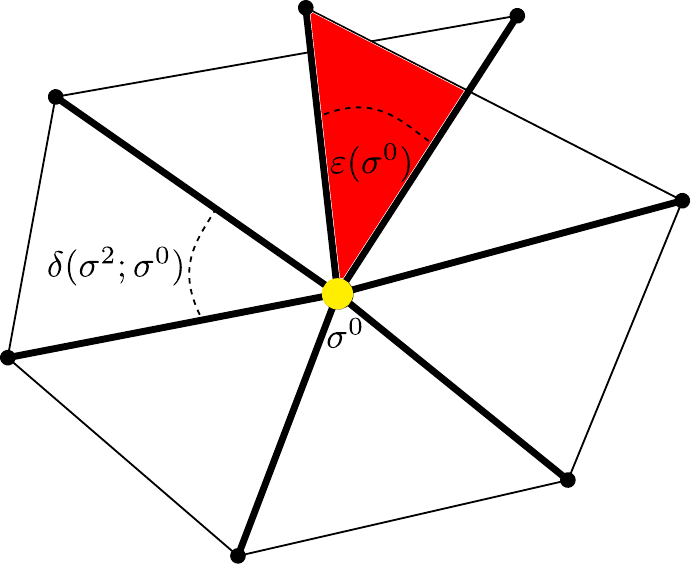}
}
\caption{A deficit angle $\varepsilon (\sigma^{0})$ is shown as it would appear when flattened on $\mathbb{R}^2$.}
\label{fig:deficit-angle}
\end{minipage}
\end{figure}

The next important operator for which there is a discrete analogue in the framework of DEC is the Hodge star $\ast: \Omega^k(T) \rightarrow \Omega^{n-k}(\star T)$. It is an isomorphism from $k$-forms to dual $(n-k)$-forms that is its own inverse up to a sign,
\begin{equation}
\ast \ast \alpha^k = (-1)^{k(n-k)}\alpha^k.
\label{eq:d-hodge-star-inverse}
\end{equation}
The explicit action of the Hodge star operator is defined with respect to the continuum wedge product $\wedge$ and the continuum inner product $|\cdot,\cdot|$ for two $k$-forms $\alpha$ and $\beta$,
\begin{equation}
\int \alpha \wedge \ast \beta=\int dV \ |\alpha,\beta|.
\label{eq:inner-product}
\end{equation}
For both sides of this relation we will make a choice for a discrete version defined through the respective action of the forms involved on the $n$-dimensional support volume $V_{\sigma^k}$ of a $k$-simplex $\sigma^k$, namely\footnote{Our calculation of the factor $\frac{V_{\sigma^k}}{|\sigma^k||\star \sigma^k|}=\frac{k!(n-k)!}{n!}$ does not agree with the factor $\frac{1}{n}$ given in \cite{Discrete Exterior Calculus}.}
\begin{equation}
\langle \alpha \wedge \ast \beta, V_{\sigma^k} \rangle= V_{\sigma^k}\frac{\langle \alpha,\sigma^k \rangle}{|\sigma^k|}\frac{\langle \ast \beta , \star \sigma^k \rangle}{|\star \sigma^k|} = \frac{k!(n-k)!}{n!}\langle \alpha,\sigma^k \rangle \langle \ast \beta , \star \sigma^k \rangle,
\label{eq:discrete-wedge}
\end{equation}
and
\begin{equation}
\langle |\alpha,\beta | dV ,V_{\sigma^k}\rangle =V_{\sigma^k} \frac{\langle \alpha,\sigma_k \rangle}{|\sigma^k|} \frac{\langle \beta, \sigma^k \rangle}{|\sigma^k|},
\label{eq:discrete-metric}
\end{equation}
where $|\sigma^k|$ is the volume of the simplex $\sigma^k$. For a vertex we choose $|\sigma^0|=1$. Equating these two expressions defines the action of the discrete Hodge star operator explicitly,
\begin{equation}
\langle \ast \alpha^k,\star\sigma^k \rangle = \frac{|\star \sigma^k|}{|\sigma^k |} \langle \alpha,\sigma^k \rangle.
\label{eq:d-hodge-star}
\end{equation}
With these definitions we obtain a discrete analogue for the integrated $L^2$-norm of two $k$-forms $\alpha^k$ and $\beta^k$,
\begin{equation}
\int dV \ |\alpha^k,\beta^k|\rightarrow\frac{k!(n-k)!}{n!}\sum_{\sigma^k}\langle\ast\alpha^k,\star\sigma^k\rangle\langle\beta^k,\sigma^k\rangle.
\label{eq:discrete-norm}
\end{equation}
With the Hodge star operator, we can define the adjoint of the exterior derivative $d$ with respect to the inner product in eq.\ \eqref{eq:discrete-norm}. The adjoint of $d$ is the discrete analogue of the codifferential $\delta :\Omega^{k+1}(T) \rightarrow \Omega^{k}(T)$ and is given by
\begin{equation}
\delta=(-1)^{nk+1}\ast d\ast.
\label{eq:d-cod-def}
\end{equation}
We can also define the discrete Ricci scalar $R(\sigma^1)$ evaluated on a link $\sigma^1$, as an average over the dual areas $|\star \sigma^0|$ of the two vertices $\sigma^0$ contained in a one-simplex $\sigma^1$,
\begin{equation}
R(\sigma^1)=\sum_{\sigma^0\prec\sigma^1}\frac{\varepsilon(\sigma^0)}{|\star \sigma^0|}.
\label{eq:gauss}
\end{equation}
Here, $\varepsilon(\sigma^0)$ is the deficit angle, which in two dimensions is defined on the vertex $\sigma^0$,
\begin{equation}
\varepsilon (\sigma^{0})=2\pi-\! \! \! \! \!\! \sum_{\sigma^{n}\succ \sigma^{(n-2)}}\!\!\!\!\delta(\sigma^2;\sigma^0),
\end{equation}
where $\delta(\sigma^2;\sigma^0)$ is the dihedral angle of the vertex $\sigma^0$ inside the triangle $\sigma^2$. For an illustration of the deficit angle around a vertex $\sigma^0$, see \fref{fig:deficit-angle}. The triangulation around the yellow vertex $\sigma^0$ is ``flattened" on $\mathbb{R}^2$. The triangles around $\sigma^0$ will either overlap (negative curvature) or will not close (positive curvature) if the deficit angle around a vertex is not equal to zero. In the triangulation in \fref{fig:deficit-angle}, the excess of the triangles around $\sigma^0$ in comparison to flat space is marked in red and the corresponding negative deficit angle $\varepsilon (\sigma^0)$ is marked with a dotted line. The dihedral angle $\delta(\sigma^2;\sigma^{0})$ for a single triangle $\sigma^2$ is also marked with a dotted line.

In two dimensions, the choice for $R(\sigma^1)$ in eq.\ \eqref{eq:gauss} is consistent with the discrete version of the Gauss-Bonnet theorem if multiplied by the support volume $V_{\sigma^1}$ of the link $\sigma^1$ and divided by $2$, i.e. 
\begin{equation}
\sum_{\sigma^1}V_{\sigma^1}\frac{R(\sigma^1)}{2}=2\pi\chi(\mathcal{M}).
\label{eq:gauss-bonnet-ricci}
\end{equation}
It was shown in \cite{Ben-Chen:2010} that these operators of DEC are suitable for defining the discrete analogue of the Killing energy $E(\omega)$ of eq.\ \eqref{eq:killing-energy-s} for a closed two-dimensional simplicial manifold $T$. We note that the operators $d$, $\delta$ and $R$ can all be represented as matrices acting on the vectors that represent the one-forms $\sigma^1$. The matrix corresponding to $d$ is of size $N_1\times N_{2}$, the matrix corresponding to $\delta$ is of size $N_1\times N_{0}$ and $R$ is a diagonal $N_1\times N_1$ matrix. The entries of these matrices for a given simplicial complex $T$ are defined by the evaluation of eqs.\ \eqref{eq:exterior-derivative}, \eqref{eq:d-hodge-star}, \eqref{eq:d-cod-def}, and \eqref{eq:gauss} on the links contained in $T$.

We now turn to the discrete Killing energy $E(\omega)$ of a discrete one-form $\omega$. By use of eq.\ \eqref{eq:discrete-norm}, it takes the form
\begin{equation}
E(\omega) = \frac{1}{2}\sum_{\sigma_1 \in T}\langle \ast S\omega, \star \sigma_1 \rangle \langle\omega,\sigma_1\rangle=\lambda.
\label{eq:killing-energy-d}
\end{equation}
The discrete operator $S$ has the exact same form as in eq.\ \eqref{eq:killing-operator} where the operators $d$, $\delta$ and $R$ are replaced by their discrete counterparts. For an easier implementation of numerical diagonalisation methods for finding the eigenvalues and eigenvectors of $S$, we will normalise $\omega$ such that the squared sum of its components is equal to one,
\begin{equation}
\sum_i (\omega_i)^2 = 1,
\label{eq:ortho-norm}
\end{equation}
instead of a normalisation in terms of the inner product in eq.\ \eqref{eq:discrete-norm}. With respect to this normalisation we can equivalently solve the eigenvalue problem for
\begin{equation}
\tilde{S}\omega=\lambda\omega, \quad \tilde{S} = \frac{1}{2}\! \ast\! S,
\label{eq:trivial-eigenvalue-problem}
\end{equation}
using the orthonormal normalisation of $\omega$ given in eq.\ \eqref{eq:ortho-norm}. The operator $\tilde{S}$ can now be represented as a matrix of dimension equal to the number of links in $T$, with entries derived from the evaluation of $\langle \tilde{S}\omega, \star \sigma^1 \rangle$. In our case of two-dimensional CDT it is a sparse matrix, which is advantageous for the numerical methods we will discuss in \sref{sec:example-geometries}.
The spectrum $\{ \lambda_i \}$ of the Killing energy operator $\tilde{S}$ is found by solving the eigenvalue problem defined by eq.\ \eqref{eq:trivial-eigenvalue-problem}. Remember that for a given topology and dimension there is a maximum number of possible Killing symmetries, implying a maximum number of eigenvectors of $\tilde{S}$ that are potential discrete analogues of the one-forms that are dual to approximate Killing vectors.

After finding a one-form $\omega$ corresponding to an eigenvalue $\lambda$ we want to be able to discuss the associated discrete vector field. We choose to use a definition of discrete vector fields $v \in \mathfrak{X}(\star T)$,
\begin{equation}
\mathfrak{X}(\star T)\equiv\bigsqcup_{\star\sigma^n\in \star T}\!\! T_{c(\sigma^n)}\sigma^n,
\end{equation}
in terms of the disjoint union of the tangent spaces $T_{c(\sigma^n)}\sigma^n$ associated to the dual vertices $\star\sigma^n \in \star T$, which are equal to the circumcenters $c(\sigma^n)\! =\! \star\sigma^n$ of the $n$-simplices $\sigma^n$. A definition in terms of the tangent space on the primal vertices would be ambiguous.

\begin{figure}
\centering
\includegraphics[width=0.7\linewidth]{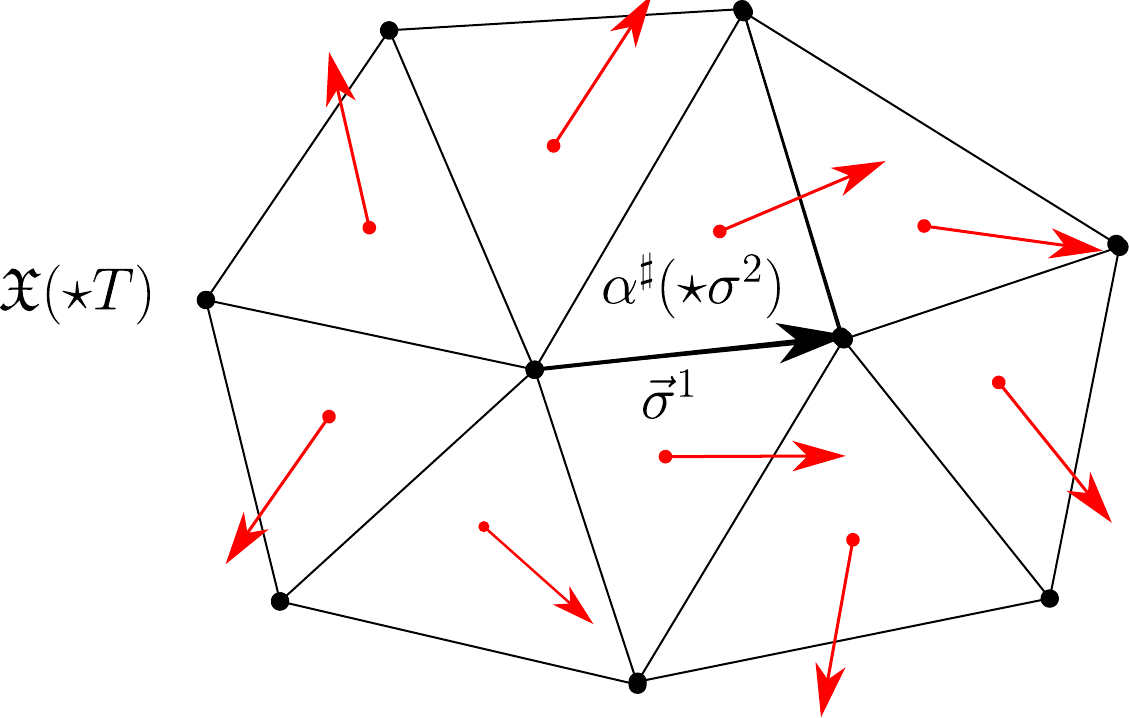}
\caption{A discrete vector field $\alpha^\sharp\in\mathfrak{X}(\star T)$ (in red) on a piece of a two-dimensional simplicial manifold $T$. The vector field is determined by evaluating the discrete one-form $\alpha^1$ on the links with link vector $\vec{\sigma}^1$. A vector $\alpha^\sharp(\star\sigma^n)$ is an element of the tangent space of the dual vertex $\star\sigma^n$. The dual vertices are shown as red dots.}
\label{fig:discrete-vector-field}
\end{figure}

To define a map from $\omega$ to its associated discrete vector field we need a discrete sharp operator $\sharp: \Omega^1(T) \rightarrow \mathfrak{X}(\star T)$. There are many choices for discrete sharp operators, with various properties \cite{Discrete Exterior Calculus}. We choose a particularly simple one,
\begin{equation}
\alpha^\sharp(\star\sigma^n)=\sum_{\sigma^1\prec\sigma^n}\langle\alpha^1,\sigma^1 \rangle \vec{\sigma}^1.
\label{eq:sharp}
\end{equation}
\fref{fig:discrete-vector-field} shows an example of a discrete vector field.
With the sharp operator $\sharp$ we can construct the dual discrete vector fields corresponding to the discrete one-forms $\omega$, which are the eigenvectors of $\ast S$. The vector field constructed from the eigenvectors corresponding to the lowest eigenvalues are called discrete approximate Killing vector fields (DAKVFs).

\section{Discrete Killing vector fields}
\label{sec:dkvf}

Finding the minimum eigenvalue of the operator $S$ in eq.\ \eqref{eq:killing-operator} is equivalent to minimising the Killing energy $E(\omega)$ given in eq.\ \eqref{eq:killing-energy}. In \sref{sec:dec} we derived the discrete counterparts of all the operators appearing in $S$ and showed that in the discrete case we can equivalently solve the eigenvalue problem for $\tilde{S}$ in eq.\ \eqref{eq:trivial-eigenvalue-problem}. The operator $\tilde{S}$ is a linear transformation on the space of discrete one-forms $\omega$. A basis for a general one-form $\omega$ can be given for a simplicial complex $T$ with $N_1$ links $\sigma^1_j$, in terms of $N_1$ basis one-forms $\omega_i$ that act as a Kronecker delta on $\sigma^1_j$, i.e.
\begin{equation}
\langle \omega_i,\sigma_j \rangle = \delta_{ij}.
\label{eq:one-form-basis}
\end{equation}
For a general one-form $\omega$ we can then write
\begin{equation}
\omega = \sum^{N_1}_{i=1} a_i \omega_i.
\label{eq:one-form}
\end{equation}
In other words, $\omega$ is an element of the vector space spanned by the $\omega_i$. A basis one-form $\omega_i$ is determined by its evaluation on the 1-simplices (links) $\sigma_j^1 \in T$. We can therefore deduce the form of $\tilde{S}$ by deriving how the one-form $\omega_i'\equiv \tilde{S}^j_{\hphantom{j}i}\omega_j$ acts on an arbitrary link $\sigma_j^1$. We will do this term by term using eqs.\ \eqref{eq:exterior-derivative} and \eqref{eq:d-cod-def}. We get 
\begin{equation}
\langle (\ast\delta d \omega)_i,\star \sigma_j^1 \rangle = \sum_{\sigma^2 \succ \sigma_i^1} \sum_{\bar{\sigma}_l^1 \prec \sigma^2}\frac{|\star \sigma^2|}{|\sigma^2|}\,\textrm{sign}(\star \sigma^2;\star \sigma_i^1) \, \textrm{sign}(\bar{\sigma}_l^1;\sigma^2) \langle \omega_l, \sigma_j^1 \rangle,
\label{eq:cod-d}
\end{equation}
and
\begin{equation}
\langle (\ast d\delta\omega)_i, \star \sigma_j^1 \rangle = \sum_{\sigma^0\prec\sigma_i^1}\sum_{\bar{\sigma}_l^1\succ\sigma^0}\frac{|\sigma^0|}{|\star\sigma^0|} \frac{|\star \sigma_i^1|}{|\sigma_i^1|}\frac{|\star \bar{\sigma}_l^1|}{|\bar{\sigma}_l^1|} \,\textrm{sign}(\sigma^0;\sigma_i^1)\,\textrm{sign}(\bar{\sigma}_l^1;\sigma^0)\langle \omega_l,\bar{\sigma}_j^1 \rangle,
\label{eq:d-cod}
\end{equation}
for the first two terms and
\begin{equation}
\langle (\ast R\omega)_i,\star \sigma_j^1 \rangle=\sum_{\sigma^0\prec\sigma_i^1}\frac{|\star \sigma_i^1|}{|\sigma_i^1|}\frac{\varepsilon(\sigma^0)}{|\star\sigma_0|}\langle \omega_i,\sigma_j^1 \rangle,
\label{eq:edge-gaussian-curvature}
\end{equation}
for the last term in $\tilde{S}$, where $\varepsilon(\sigma^0)$ is the deficit angle. 

\begin{figure}
\centering
\begin{subfigure}{0.22\textwidth}
\includegraphics[width=\linewidth]{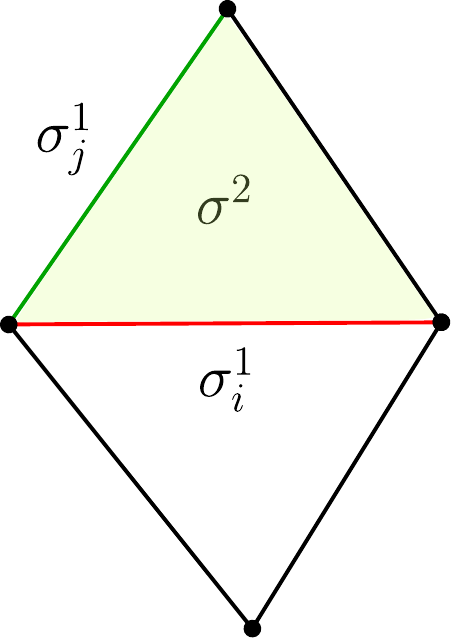}
\end{subfigure}
\begin{subfigure}{0.35\textwidth}
\includegraphics[width=\linewidth]{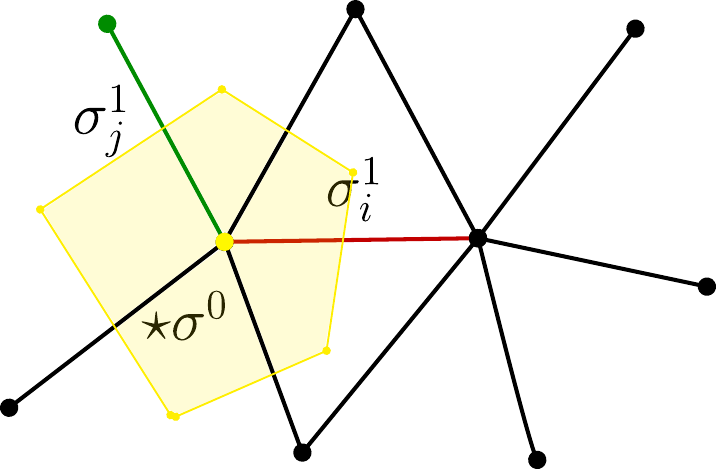}
\end{subfigure}
\begin{subfigure}{0.38\textwidth}
\includegraphics[width=\linewidth]{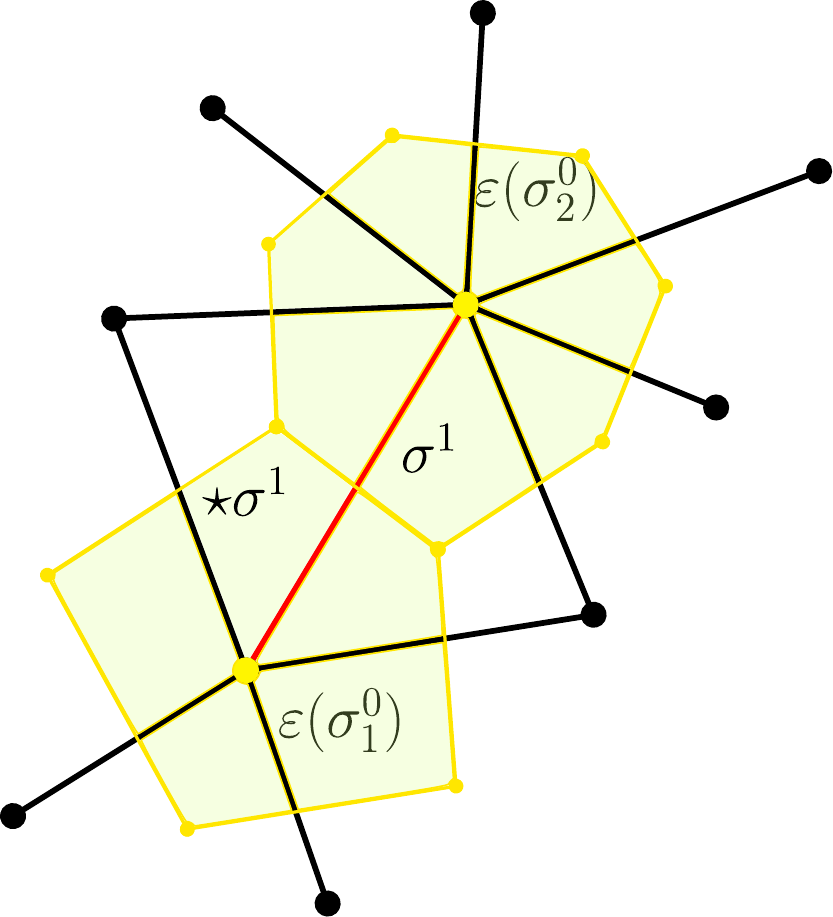}
\end{subfigure}
\caption{The geometric objects relevant to (from left to right) $\delta d$, $d \delta$ and $R(\sigma^1)$ are illustrated here. These include the link $\sigma^1_i$ (red), the link $\sigma^1_j$ (green), the triangle $\sigma^2$ (light green) and the dual area $\star\sigma^2$ (light yellow) to $\sigma^0$. The rightmost figure which illustrates $R(\sigma^1)$ also includes the deficit angles $\varepsilon{\sigma_1^0}$ and $\varepsilon{\sigma_2^0}$ of the two vertices $\sigma_1^0$ and $\sigma_2^0$ and the dual link $\star \sigma^1$ (dark yellow) orthogonal to $\sigma^1$ (red). \label{fig:gauss-neighbors}}
\end{figure}

We see that the evaluation of $\langle \tilde{S}\omega_i, \star\sigma_j^1 \rangle$ is given by a sum over the action of the basis one-forms $\omega_l$ on the links $\sigma^1_j$. For $(\delta d)^i_{\hphantom{i}j}$ these are the links $\sigma_i$ and $\sigma_j$ that share a triangle $\sigma^2$, for $(d \delta)^i_{\hphantom{i}j}$ these are the links  $\sigma_i$ and $\sigma_j$ that share a vertex. The geometric quantities that appear in the coefficients of $(\delta d)^i_{\hphantom{i}j}$, $(d \delta)^i_{\hphantom{i}j}$ and $R(\sigma^1)^i_{\hphantom{i}j}$ are illustrated in \fref{fig:gauss-neighbors}.

Eqs.\ \eqref{eq:cod-d}-\eqref{eq:edge-gaussian-curvature} show that $\tilde{S}$ is a linear transformation of $\omega$. The operator $\tilde{S}$ can be represented as a square matrix of dimension $N_1$. The entries of $\tilde{S}^i_{\hphantom{i}j}$ are given by the coefficients for $(\ast \delta d)^i_{\hphantom{i}j}$, $(\ast d \delta)^i_{\hphantom{i}j}$ and $\ast R^i_{\hphantom{i}j}$. The values and signs of the coefficients depend only on geometric properties of the neighbourhoods of the links in $T$. To be more precise, the coefficients only depend on the volumes $|\sigma^k|$ and $|\star \sigma^k|$ of the complex $T$ and the dual complex $\star T$ respectively and their relative orientation. This means that the coefficients for the triangulations considered in DT and CDT that are built from a fixed type of simplex are easily calculated. For an equilateral triangulation the contributions to the matrix coefficients are given by
\begin{equation}
\frac{|\star \sigma^2|}{|\sigma^2|} = \frac{	4}{\sqrt{3}a^2}
\label{eq:cod-d-prefactor-equilateral}
\end{equation}
and
\begin{equation}
\frac{| \sigma^0|}{|\star \sigma^0|}\frac{|\star \sigma^1|}{| \sigma^1|}\frac{|\star \bar{\sigma}^1|}{|\bar{\sigma}^1|} = \frac{	4}{\sqrt{3}a^2N_2 (\sigma^0) },
\label{eq:d-cod-prefactor-equilateral}
\end{equation}
where $N_2(\sigma^0)$ is the number of triangles that contain the vertex $\sigma_0$. The value of the deficit angle averaged over a link $\sigma^1$ can also be expressed in terms of $N_2(\sigma^0)$. We find
\begin{equation}
\frac{1}{2}\sum_{\sigma^0\prec\sigma^1}\frac{\varepsilon(\sigma^0)}{|\star\sigma_0|}= 4 \pi \left(- \frac{1}{3}+ \sum_{\sigma^0\prec\sigma^1}\frac{1}{N_2(\sigma^0)} \right).
\label{eq:gauss-prefactor}
\end{equation}

\section{Example Geometries}
\label{sec:example-geometries}

As a preparation to the discussion of discrete approximate Killing vector fields (DAKVFs) on an ensemble of DT and CDT geometries, we have investigated simplicial manifolds that approximate continuum spaces with exact symmetries. We will first consider a Delaunay triangulation of a regular sprinkling of the two-sphere with a lower bound on the distance between vertices. The two-sphere is maximally symmetric and therefore admits three linearly independent Killing vector fields in the continuum. We investigate what properties the spectrum and the eigenvectors of the Killing energy operator $S$ have for simplicial manifolds which are close to the two-sphere, similar to what was done in \cite{Ben-Chen:2010}. Next, we will discuss triangulations of the flat two-torus. The continuum two-torus can be triangulated exactly and admits two Killing vector fields. The triangulation of the torus will therefore also have two exact discrete Killing vector fields. This makes these toroidal simplicial geometries a good playground for studying the behaviour of the DAKVFs under an explicit breaking of the symmetries of the torus.

\subsection{Discrete Sphere}
\label{ssec:discrete-sphere}

\begin{figure}
\centering
\includegraphics[width=0.60\linewidth]{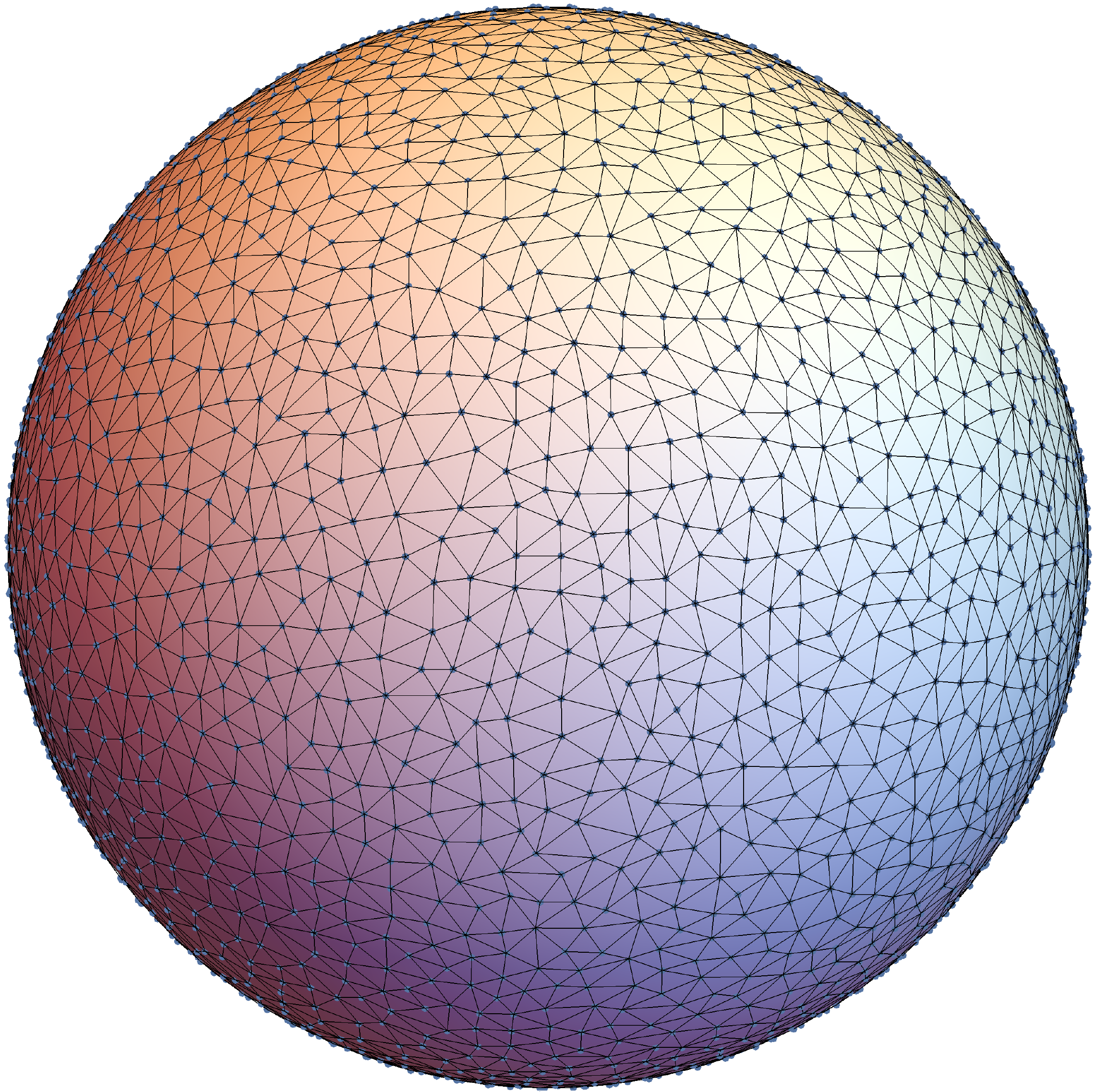}
\caption{An example of the Delaunay triangulation of a regular sprinkling of the unit two-sphere. The triangulation consists of $N_0=3086$ vertices, $N_1=9252$ links and $N_2=6168$ triangles.}
\label{fig:spherical-triangulation}
\end{figure}

The spherical simplicial geometries are generated by a Delaunay triangulation of a regular sprinkling of the two-sphere of radius $r$ with a lower bound on the distance between vertices. The process of sprinkling will not be described further here. In a Delaunay triangulation of the two-sphere, no vertex lies within the circumcircle of any triangle. This property ensures an upper bound on the link length in the triangulation. As a consequence, the resulting triangulations are relatively close to equilateral. An example of a Delaunay triangulation is given in \fref{fig:spherical-triangulation}.

\begin{figure}
\centering
\includegraphics[width=0.80\linewidth]{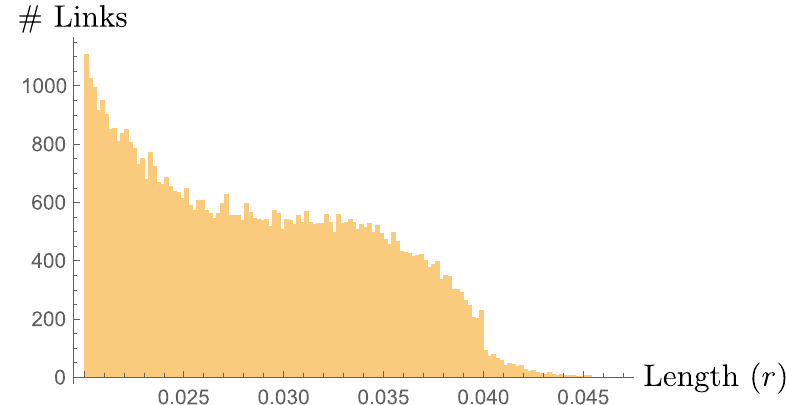}
\caption{The distribution of links of length in terms of the length scale $r$ for a spherical Delaunay triangulation with a total of $N_1=57279$ links and a minimal link length of $0.02r$. The distribution is not Gaussian. However, the distribution is still relatively narrow because a Delaunay triangulation puts an upper bound on the maximum link length.}
\label{fig:link-distribution}
\end{figure}

The triangulation of the regular sprinkling is realised with a minimum link length set by hand. The resulting distribution of the link lengths for a typical triangulation is shown in \fref{fig:link-distribution}.

We are interested in studying the properties of DAKVFs on this simplicial complex to understand how well the DAKVFs encode the approximate symmetries of the geometry. Using the framework of discrete exterior calculus as described in \sref{sec:dec}, we define the discrete Killing energy operator $\tilde{S}$ on this geometry. $\tilde{S}$ takes the form of a square matrix of size $N_1$, the number of links in the geometry. The matrix is given explicitly by eqs.\ \eqref{eq:cod-d}, \eqref{eq:d-cod-def} and \eqref{eq:gauss}. The spectrum of $\tilde{S}$ is invariant under a change of orientation of the simplices in the simplicial geometry. We will make a particular simple choice. We choose an orientation based on a labeling of the vertices where we define the orientation of all higher dimensional simplices and their dual simplices with respect to an ordering of the vertex labels.

We will compare the spectrum of $\tilde{S}$ for a triangulated two-sphere with the spectrum of $S$ for a smooth two-sphere. The spectrum for a smooth two-sphere is easily calculated using a Hodge decomposition. A solution for the eigenvector $\omega$ of $S$ can be found by decomposing it into a scalar $\phi$, a two-form $\psi$ and a harmonic component $h$,
\begin{equation}
\omega = d\phi+\delta\psi+h.
\label{eq:hodge-decomposition}
\end{equation}   
By definition, the exterior derivative and the co-differential of the harmonic component vanishes, i.e. $\delta h=0$, $dh=0$. The harmonic component $h$ is zero on the sphere due to the Poincar\'e-Hopf theorem. The potentials $\phi$ and $\psi$ can be found by solving
\begin{equation}
\Delta\phi = \delta \omega,
\label{eq:scalar-hodge}
\end{equation}
and 
\begin{equation}
\Delta\psi = d \omega,
\label{eq:dual-scalar-hodge}
\end{equation}
with respect to the Laplace-Beltrami operator $\Delta=d\delta+\delta d$ on $k$-forms. In a Hodge decomposition, the eigenvalue equation for $\omega$ in eq.\ \eqref{eq:eigenvalue-problem} splits into two separate equations,
\begin{eqnarray}
4\Delta\phi-(\frac{4}{r^2}+\lambda_\phi)\phi&=&0, \notag \\
2\Delta\psi-(\frac{4 }{r^2}+\lambda_\psi)\psi&=&0,
\label{CEigen}
\end{eqnarray}
where we have used that $\phi$ and $\psi$ are only defined up to a constant by eq.\ \eqref{eq:hodge-decomposition}. The solutions for $\phi$ and $\psi$ in eq.\ \eqref{CEigen} are given by spherical harmonics. The spectrum of $S$ can therefore be given in terms of two sets of spherical harmonics for $\phi$ and $\psi$, with $\phi=0$ if $\psi\neq 0$, $\lambda=\lambda_\psi$ and $\psi=0$ if $\phi\neq 0$, $\lambda=\lambda_\phi$, and the special case $\phi\neq 0$, $\psi\neq 0$, $\lambda_\phi=\lambda_\psi=\lambda$. The set of corresponding eigenvalues is given by
\begin{equation}
\lambda \in \lbrace 4\rho_{k_\phi}-\frac{4}{r^2}, 2\rho_{k_\psi}-\frac{4}{r^2} \rbrace
\end{equation}
in terms of the two sets $\rho_\phi$ and $\rho_\psi$,
\begin{equation}
\rho_{k_\phi}=\frac{k_\phi(k_\phi+1)}{r^2}, \ \ \ \ \rho_{k_\psi}=\frac{k_\psi(k_\psi+1)}{r^2}, \ \ \ \ k_\phi,k_\psi \in \mathbb{N}_{>0}.
\end{equation}
\begin{figure}[t!]
\centering
\begin{subfigure}{0.48\textwidth}
\includegraphics[width=\linewidth]{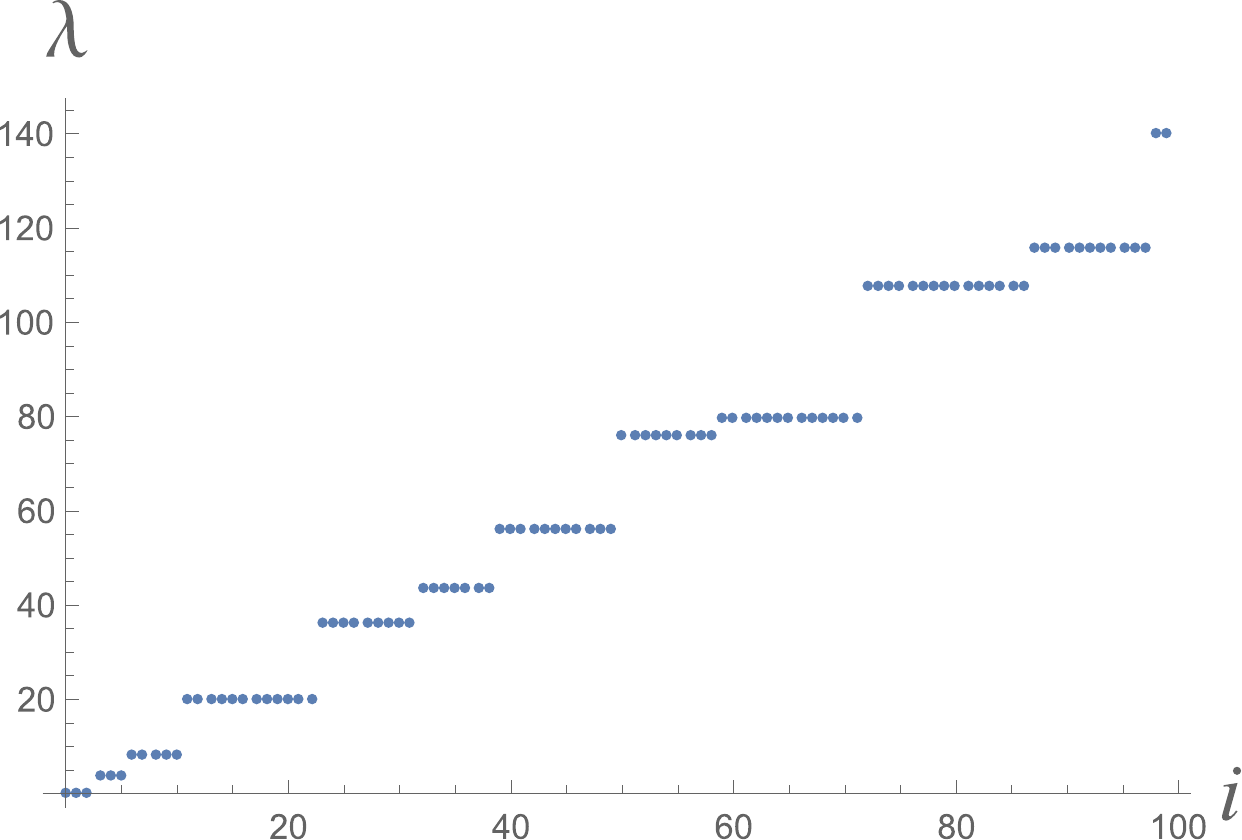}
\end{subfigure}
\hspace*{\fill}
\begin{subfigure}{0.48\textwidth}
\includegraphics[width=\linewidth]{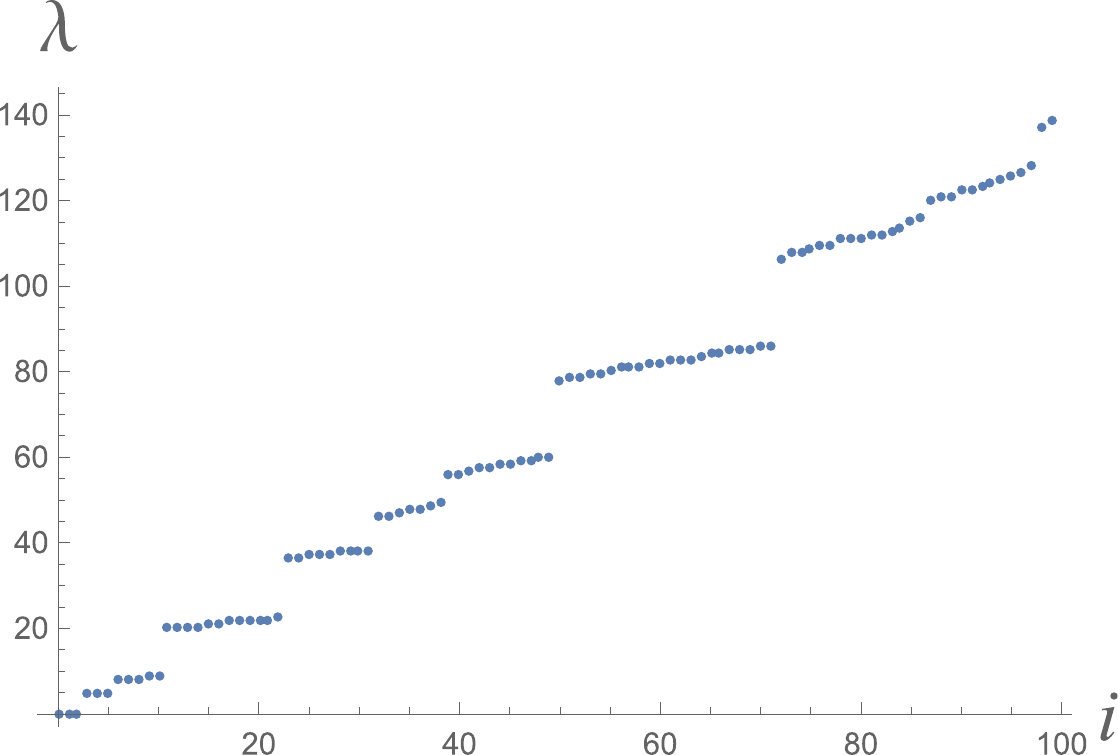}
\end{subfigure}
\caption{The lowest 100 eigenvalues $\lambda_i$ (labelled on the horizontal axis) of the Killing energy operator $S$ on the smooth two-sphere (left) and the eigenvalues of $\tilde{S}$ on a triangulation of the two-sphere (right).}
\label{fig:sphere-killing-eigenvalues}
\end{figure}
The special case $\phi\neq 0$, $\psi\neq 0$ corresponds to $k_\phi=2$ and $k_\psi=3$. The eigenvalues for either $k=k_\phi$ or $k=k_\psi$ have a degeneracy of $2k+1$ except for the special case $\phi\neq 0$, $\psi\neq 0$, which has a degeneracy of $11$. To obtain these eigenvalues, we have used that the curvature scalar on $S^2$ is $R=\frac{2}{r^2}$, where $r$ is the radius of the two-sphere. The degeneracy of the eigenvalues implies that there are three zero eigenvalues $\lambda=0$, which correspond to the three linearly independent Killing vectors on the smooth two-sphere. We can now compare this to the spectrum of the discrete analogue of the Killing energy $\tilde{S}$, which we can calculate numerically. In \fref{fig:sphere-killing-eigenvalues}, a comparison is given of the spectrum of a triangulation of the two-sphere and the spectrum of the smooth two-sphere.

\begin{figure}
\captionsetup[subfigure]{labelformat=empty}
\centering

\begin{subfigure}[t]{0.41\textwidth}
\includegraphics[width=\linewidth]{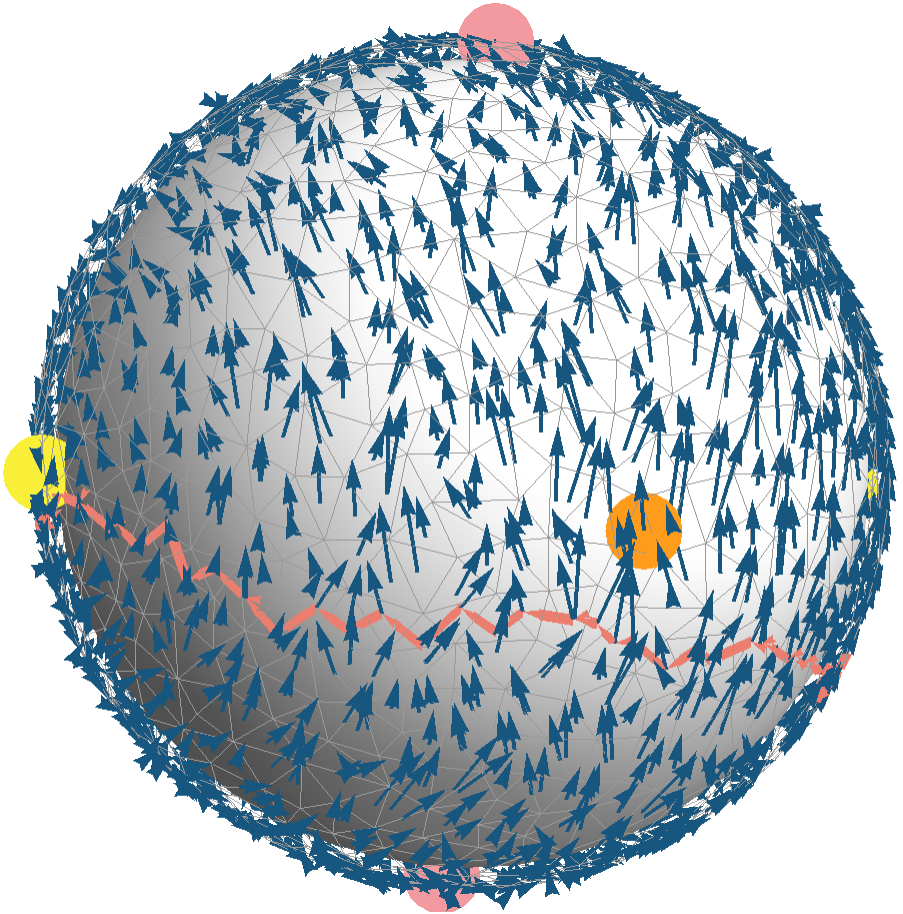}
\caption{\scalebox{2}{$\omega_1^\sharp$}}
\end{subfigure}   \hspace{4em}
\begin{subfigure}[t]{0.41\textwidth}
\includegraphics[width=\linewidth]{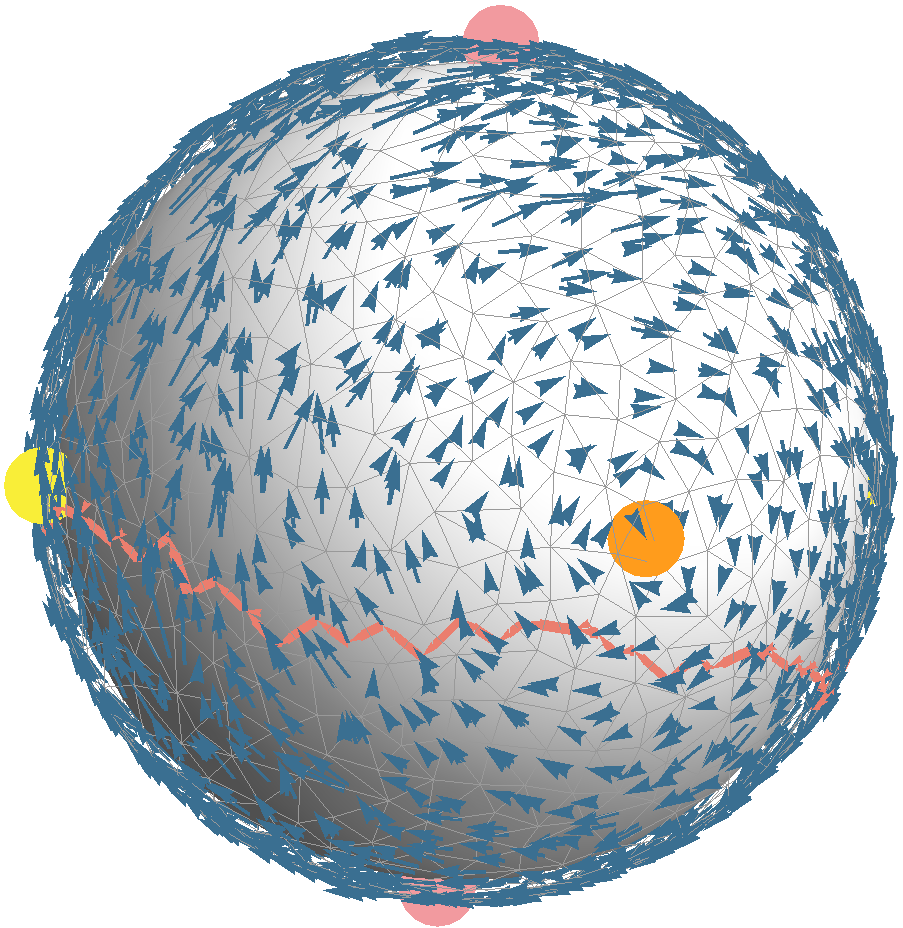}
\caption{\scalebox{2}{$\omega_2^\sharp$}}
\end{subfigure}\\ \vspace{1em}
\begin{subfigure}[t]{0.41\textwidth}
\includegraphics[width=\linewidth]{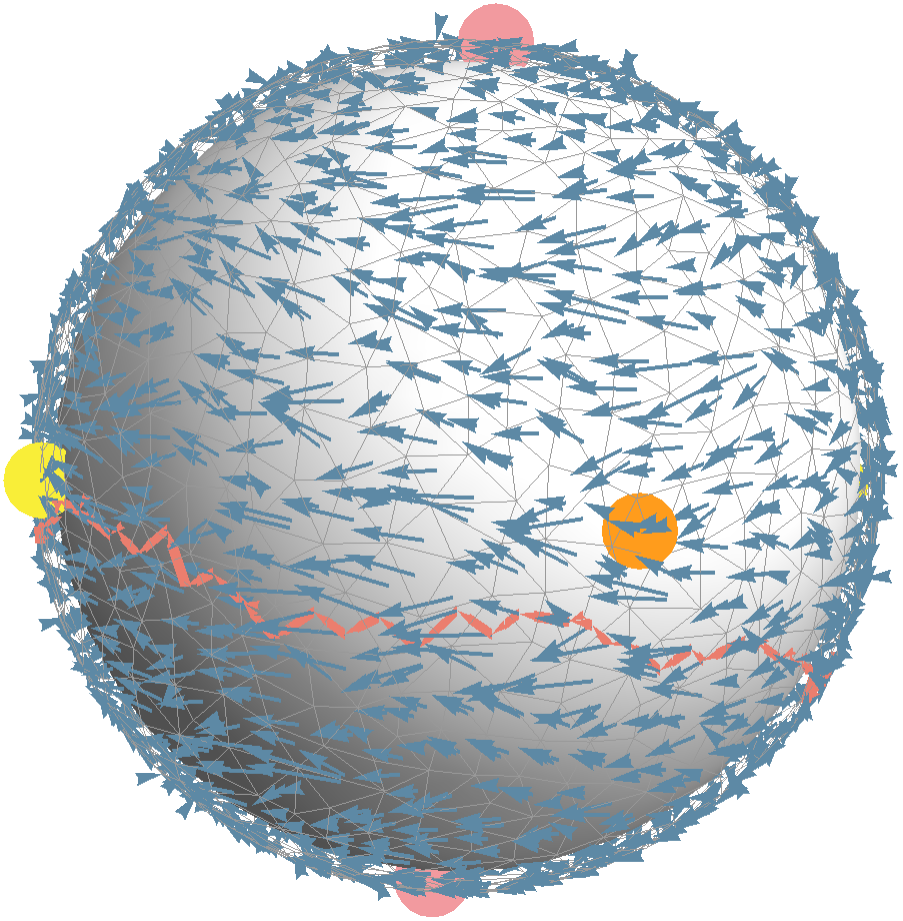}
\caption{\scalebox{2}{$\omega_3^\sharp$}}
\end{subfigure}  \hspace{4em}
\begin{subfigure}[t]{0.41\textwidth}
\includegraphics[width=\linewidth]{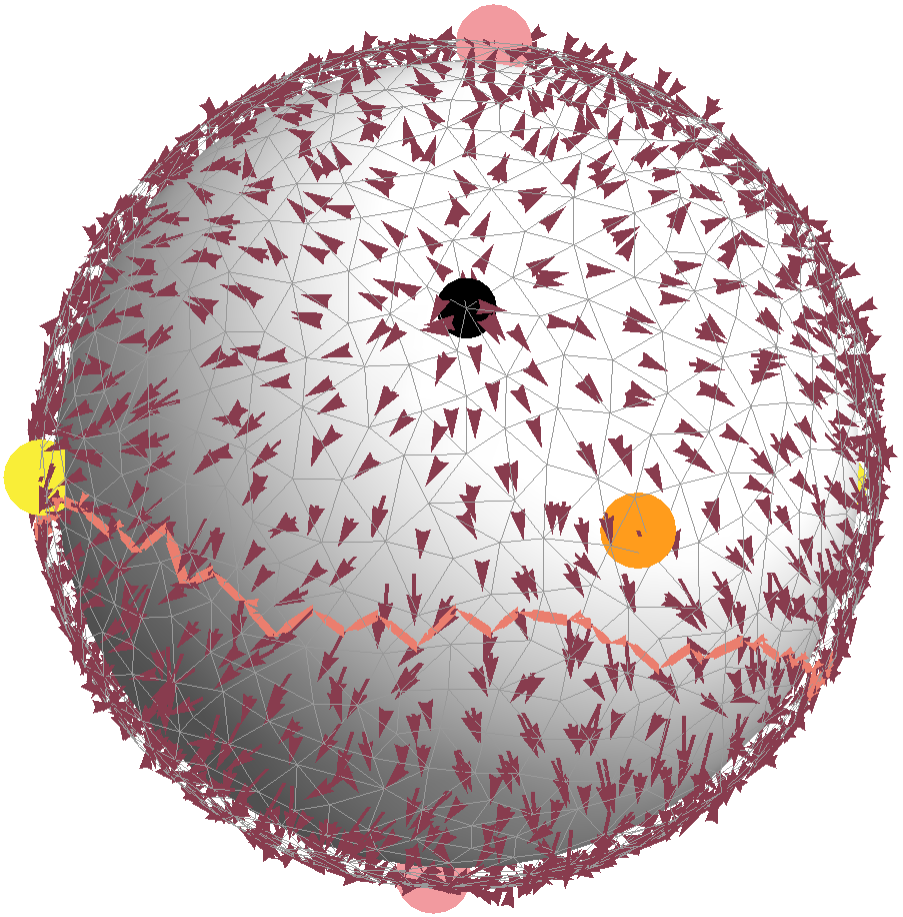}
\caption{\scalebox{2}{$\omega_4^\sharp$}}
\end{subfigure}
\caption{The vector fields associated to the four lowest eigenvalues of $\tilde{S}$ for a typical triangulated two-sphere. The vector fields $\omega_1^\sharp$, $\omega_2^\sharp$ and $\omega_3^\sharp$ (with increasingly lighter shades of blue) approximately generate a rotation around the axis that goes through the poles marked by the colored points. The purple vector field $\omega_4^\sharp$ is clearly different. It has one source (indicated by a black point), one sink, and no significant rotational component. The pink line in every figure indicates the discrete flow generated by $\omega_3^\sharp$, starting from a triangle near the equator with respect to the pole.}
\label{fig:sphere-dakvf}
\end{figure}
We see that the spectra are most similar for the lowest eigenvalues. This is to be expected, because the higher modes probe smaller scales and are therefore more sensitive to the discretisation. The lowest three eigenvalues of the spectrum on the discrete sphere are close to zero. This signals that the three associated eigenvectors are related to approximate symmetries of the two-sphere. To corroborate this statement, we can use the discrete sharp operator of eq.\ \eqref{eq:sharp} to investigate the corresponding discrete vector fields. The discrete vector field $\omega_i^\sharp$ corresponds to the $i$-th lowest eigenvalue of $\tilde{S}$.

\fref{fig:sphere-dakvf} shows the four discrete vector fields associated to the lowest four eigenvalues. For illustrative purposes we have chosen a triangulation with relatively few triangles. The properties of triangulations of the sphere with more triangles are qualitatively the same. On inspection, the three vector fields $\omega^\sharp_1$, $\omega^\sharp_2$ and $\omega^\sharp_3$, corresponding to the lowest three eigenvalues, are similar to the Killing vector fields on the smooth two-sphere. On the smooth two-sphere, these vector fields generate the three linearly independent rotations.

A thorough analysis of the convergence properties of the DAKVFs to exact Killing symmetries will not be given here. We have, however, tested their relation to exact Killing symmetries in several ways. For example, we have considered the flow along the discrete vector field on the piecewise flat triangulations. The constant flow within a two-simplex $\sigma^2_1$ is generated by the vector field obtained with the sharp operator $\sharp$ starting from the center $\star \sigma^2_1$. If the flow reaches the boundary between two two-simplices $\sigma^2_1$ and $\sigma^2_2$, the flow in the second two-simplex $\sigma^2_2$ is continued from the center $\star \sigma^2_2$ with respect to the discrete vector field at $\star \sigma^2_2$.

In this way, the discrete flow can be represented as a connected sequence of two-simplices. The jump from $\star \sigma^2_1$ to $\star \sigma^2_2$ is an additional discretisaton step. We expect that any error introduced by this discretisation of the generated flow is negligible in the limit of the number of two-simplices $N_2 \rightarrow \infty$. The pink flow lines in \fref{fig:sphere-dakvf} are an example of the discrete flows generated by these discrete vector fields. We have calculated the deviation of the flow of the discrete vector fields from the flow of their smooth counterparts and found that it is small in comparison to the length scale $\sqrt{\bar{A}_{\sigma^2} N_2}$, where $\bar{A}_{\sigma^2}$ is the average of the area of the triangles in the triangulation. This deviation also decreases with increasing $N_2$.
\begin{figure}
\captionsetup[subfigure]{labelformat=empty}
\centering
\begin{subfigure}[t]{0.41\textwidth}
\includegraphics[width=\linewidth]{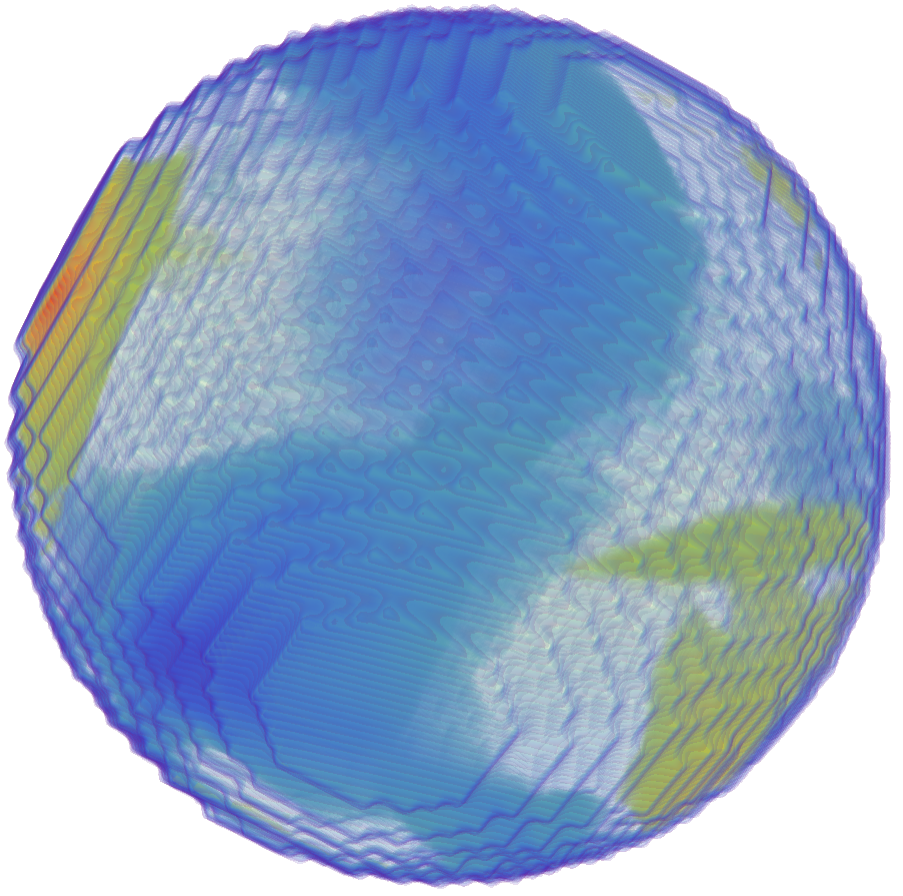}
\caption{\scalebox{2}{$\phi$}}
\end{subfigure}   \hspace{4em}
\begin{subfigure}[t]{0.41\textwidth}
\includegraphics[width=\linewidth]{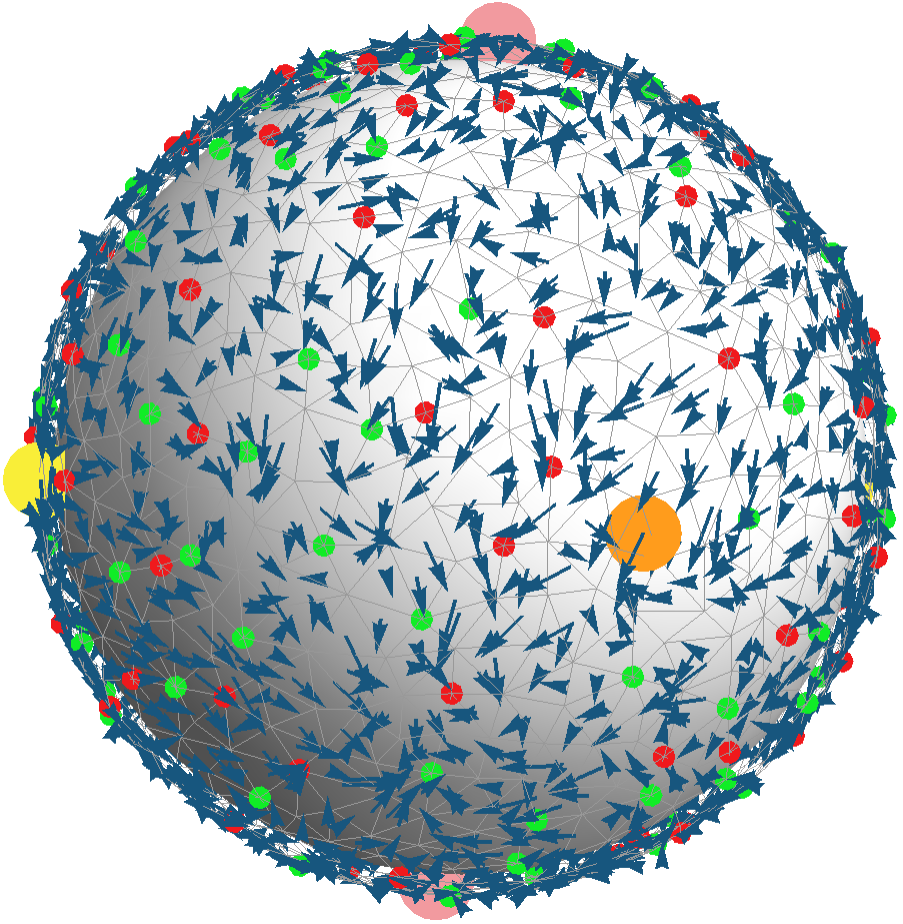}
\caption{\scalebox{2}{$d\phi$}}
\end{subfigure}\\ \vspace{1em}
\begin{subfigure}[t]{0.41\textwidth}
\includegraphics[width=\linewidth]{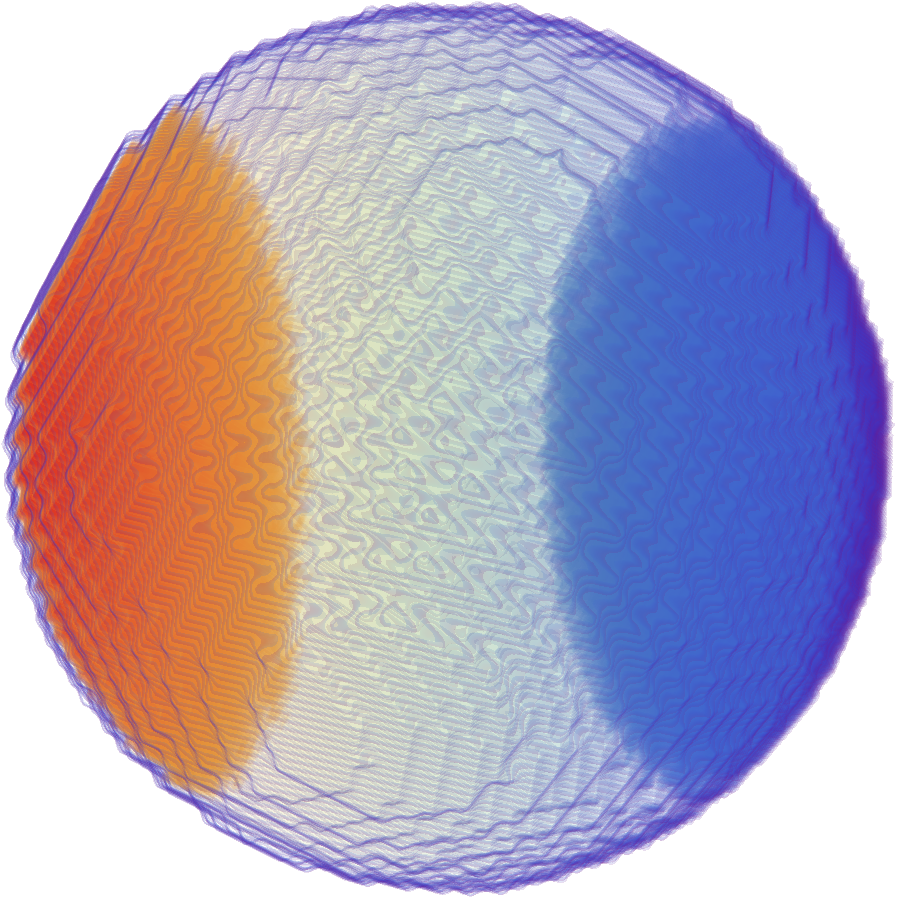}
\caption{\scalebox{2}{$\psi$}}
\end{subfigure}  \hspace{4em}
\begin{subfigure}[t]{0.41\textwidth}
\includegraphics[width=\linewidth]{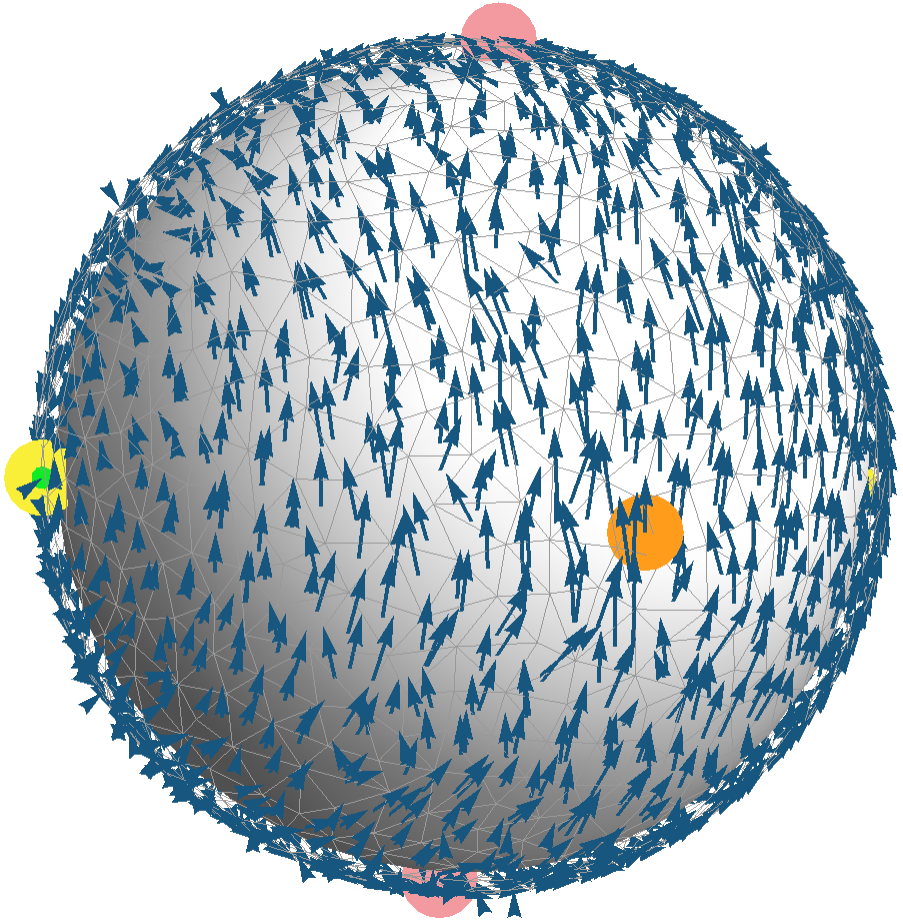}
\caption{\scalebox{2}{$\delta \psi$}}
\end{subfigure}
\caption{Values of the scalar potential $\phi$ and the dual scalar potential $\psi$ for a typical triangulated two-sphere are shown on the left. Orange corresponds to large values and blue to small values of $\phi$ and $\psi$. The decomposition of a DAKVF on a triangulated two-sphere $S^2$ into a divergence component $d\phi$ and a rotational component $\delta\psi$ is shown on the right. The green dots on the right are local maxima of $\phi$ and $\psi$ and the red dots are local minima of $\phi$ and $\psi$. In the bottom right figure, the global maximum and minimum are seen to coincide with the vortices of the vector field. The pink, yellow and orange dots are the same poles as shown in \fref{fig:sphere-dakvf}, to act as a reference. From the integrated norms $|d \omega| \approx 0.25$ and $|\delta \omega| \approx 0.96$, we observe that the rotational component is dominant for this DAKVF.}
\label{fig:decomposition-field}
\end{figure}

An additional relevant observation is the behaviour of the discrete vector field $\omega^\sharp_4$, which corresponds to the fourth-lowest eigenvalue of $\tilde{S}$. The discrete vector field $\omega_4^\sharp$ is clearly different. It has one source and one sink and does not have a strong rotational component. We have repeated the investigations illustrated in \fref{fig:sphere-dakvf} for many ($\approx 100$) different triangulations of the two-sphere and have seen the same qualitative results. We conclude that the eigenvectors of $\tilde{S}$ associated with the lowest three eigenvalues are most likely indicative of a discrete analogue of the three Killing vectors on the sphere.

To analyse the properties of the Killing vector fields in a more quantitative manner, we can make use of the discrete analogue of the Hodge decomposition in eq.\ \eqref{eq:hodge-decomposition}. We can find a discrete potential $\phi$ and discrete dual potential $\psi$ of the divergence and rotational part of the discrete one-form $\omega_i$ respectively. We find $\phi$, $\psi$ and $h$ by solving the discrete analogues of eqs.\ \eqref{eq:scalar-hodge} and \eqref{eq:dual-scalar-hodge}. Consistently comparing the two potentials for the three Killing vectors $\omega^\sharp_1$, $\omega^\sharp_2$ and $\omega^\sharp_3$ and the next eigenvector $\omega^\sharp_4$ shows that the rotational part $\delta \psi$ dominates the divergence component $d \phi$ for the three lowest eigenvalues, while the opposite is true for the fourth-lowest eigenvalue.

The vector fields dual to the two components $\delta \psi$ and $d \phi$ of the discrete one-form $\omega$ can also be considered and they are consistent with the conclusion that the three lowest eigenvectors of $\tilde{S}$ have properties very similar to those of the exact Killing vector fields of $S^2$. \fref{fig:decomposition-field} shows the potential and vector fields of a Hodge decomposition of the discrete one-form $\omega_1$ of \fref{fig:sphere-dakvf} associated to a DAKVF on $S^2$. The harmonic component $h$ of a smooth one-form $\omega$ is always zero on a two-sphere, which within numerical precision is also true for the discrete one-forms.

Although the potentials $\phi$ and $\psi$ are interesting for studying the properties of the discrete approximate Killing vector fields, they are rather coarse tools. Even these very regular spherical triangulations have a relatively large divergence component $\phi$ in the approximate Killing vector field as can be seen in \fref{fig:decomposition-field}. It seems that this definition of the divergence and rotational component of a discrete vector field is very sensitive to the details of the discretisation, as can be seen from the many maxima and minima of $\phi$ in \fref{fig:decomposition-field}.

For much less regular triangulations, like those appearing in CDT, it will be even harder to interpret the results of the discrete Hodge decomposition. Unless one can find a coarse-graining procedure to produce more robust outcomes, we expect that the discrete Hodge decomposition we have presented here is most likely not suitable for constructing quantum observables. In \sref{sec:symmetries-qg} we will introduce and study a new observable, which is also based on discrete vector fields, but has potentially more promising properties.
\begin{figure}[!t]
\centering
\includegraphics[width=0.7\linewidth]{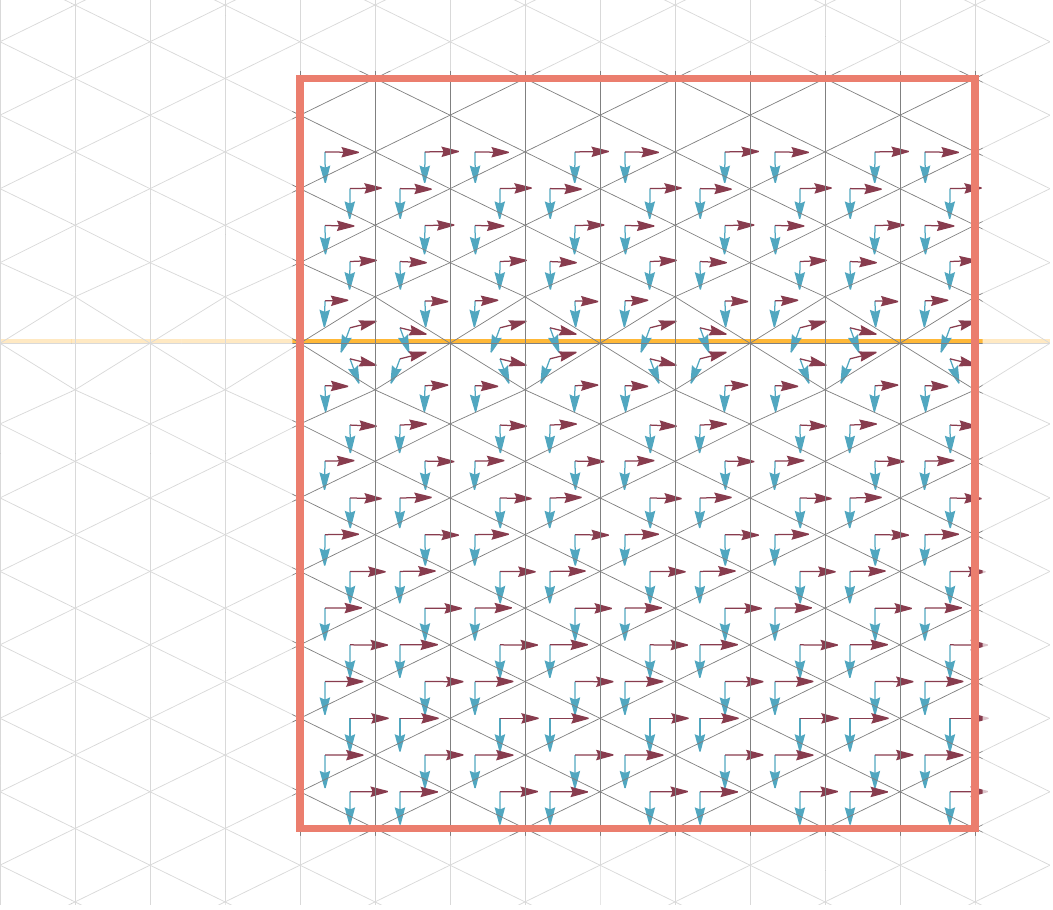}
\caption{An illustration of the behaviour of DAKVFs under symmetry breaking. The triangulation repeats periodically outside the border, which denotes the fundamental domain of the torus with a line defect. The horizontal purple vector field {\color{cpurple}$\rightarrow$} corresponds to the lowest eigenvalue of $\bar{S}$. The vertical blue vector field {\color{cblue}$ \downarrow$} corresponds to the second-lowest eigenvalue of $\bar{S}$. 
The Killing energy of the blue vector field is therefore larger than the Killing energy of the purple vector field. We conclude that the original symmetry orthogonal to the line defect is broken more strongly than the symmetry in the other directions.
}
\label{fig:broken-symmetry}
\end{figure}
\subsection{Broken symmetries on the discrete torus}
\label{ssec:discrete-torus}

Another simple two-dimensional geometry that can be studied is the flat two-torus $T^2$. The torus admits at most two Killing vectors. We will consider exact triangulations of the flat torus in terms of equilateral triangles. In this case the triangulation has two exact Killing vectors even in the piecewise flat context. This provides an ideal set-up to study how the Killing energy and DAKVFs behave under an explicit breaking of the two exact Killing symmetries that are present on the two-torus.

To discuss symmetry breaking on the two-torus, we introduce a line defect on a flat torus by hand. Along this line, the discrete curvature alternates between positive (vertex order four) and negative (vertex order eight), while remaining zero (vertex order six) everywhere else. We have then constructed the two DAKVFs that correspond to the two eigenvectors $\omega$ of $\tilde{S}$ with the two lowest eigenvalues. By introducing these line defects, we expect to break the symmetry in the direction orthogonal to the line defect stronger than the symmetry in the direction parallel to the line. This is visible in the two DAKVFs that we find. The torus with a line defect and the two DAKVFs are illustrated in \fref{fig:broken-symmetry}.

The DAKVF with the higher Killing energy (the blue vector field) generates a flow orthogonal to the line defect (the orange line). The DAKVF with the lower Killing energy (the purple vector field) generates a flow parallel to the line defect. This ordering of the DAKVFs is what one would have expected from the introduction of a line defect, which suggests that an observable based on the properties of the discrete vector fields could potentially measure how strongly the symmetries of the torus are broken.

\begin{figure}[t!]
\centering
\begin{subfigure}{.25\textwidth}
  \centering
  \includegraphics[width=\linewidth, height=6.5cm]{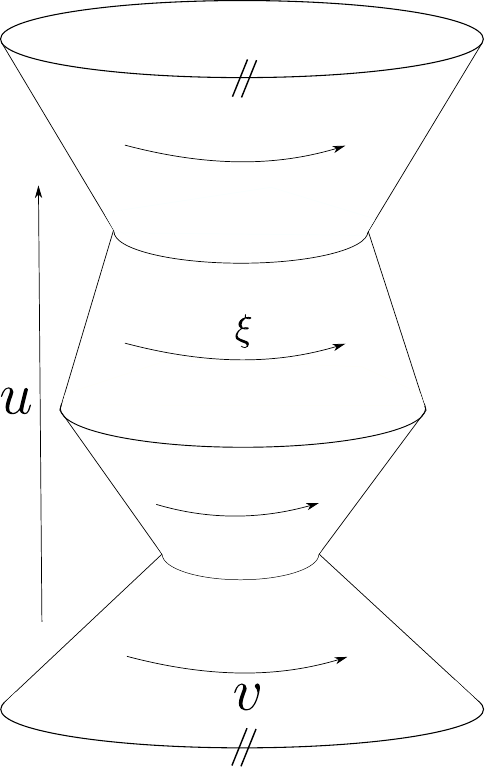}
\end{subfigure}%
\begin{subfigure}{.25\textwidth}
  \centering
  \includegraphics[width=\linewidth, height=6.5cm]{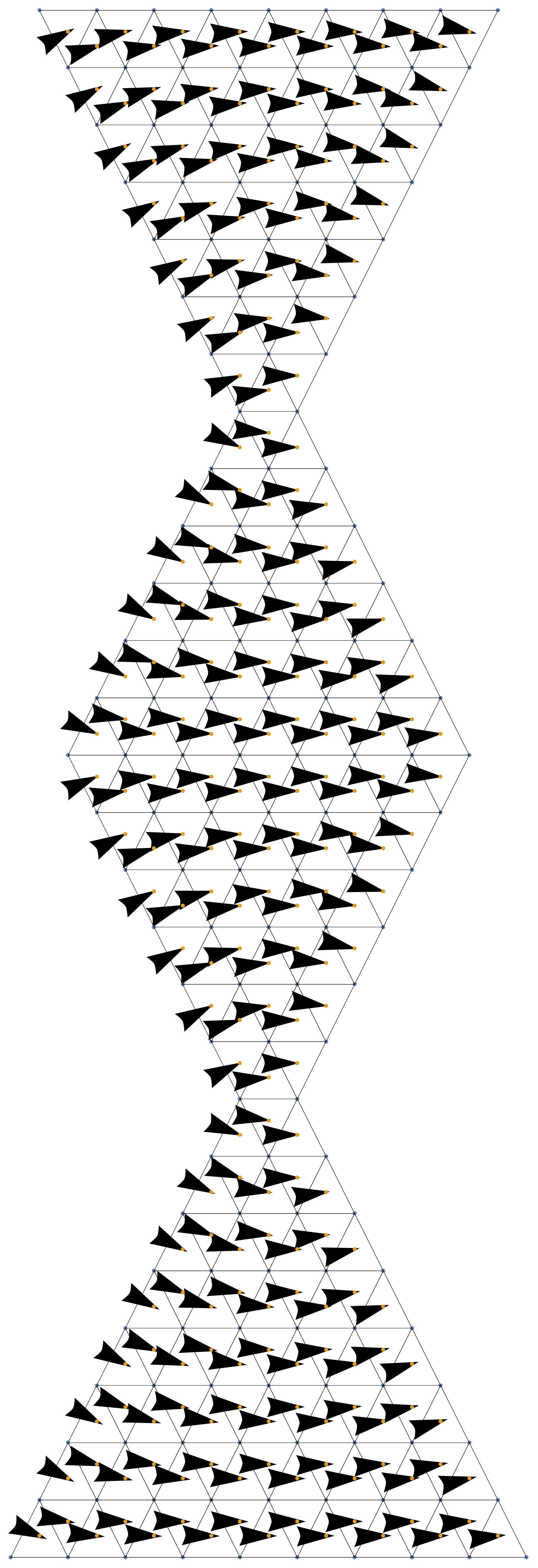}
\end{subfigure}%
\begin{subfigure}{.25\textwidth}
  \centering
  \includegraphics[width=\linewidth, height=6.5cm]{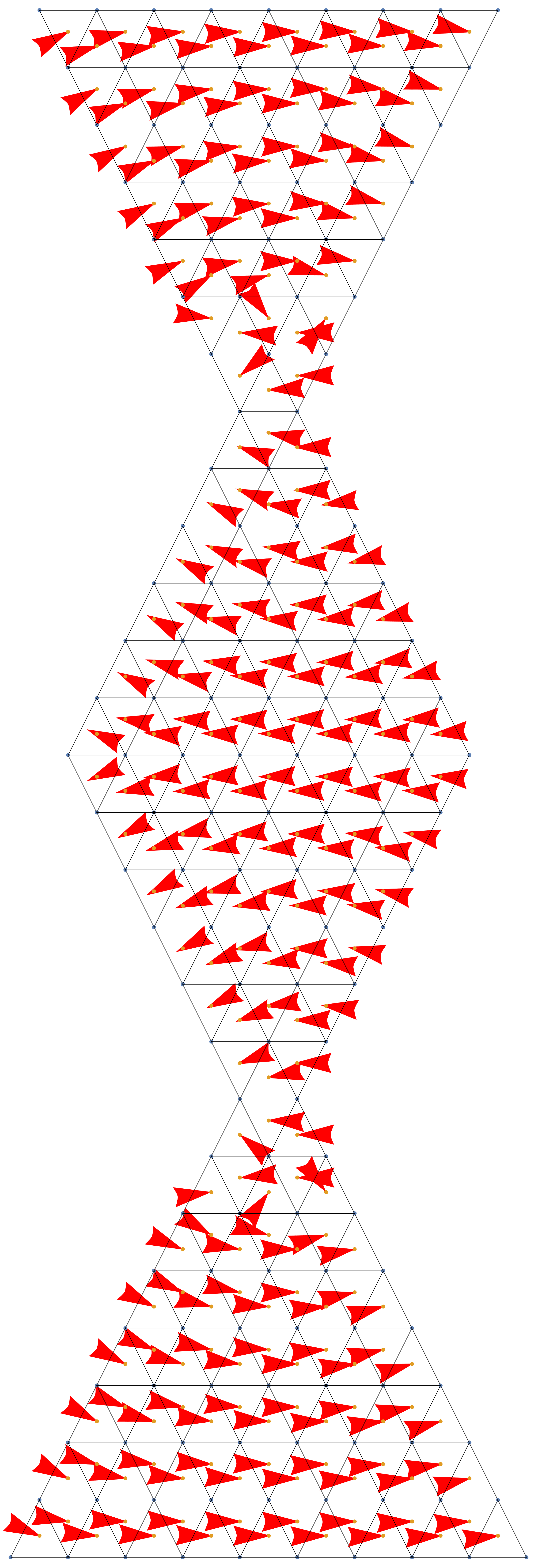}
\end{subfigure}%
\begin{subfigure}{.25\textwidth}
  \centering
  \includegraphics[width=\linewidth, height=6.5cm]{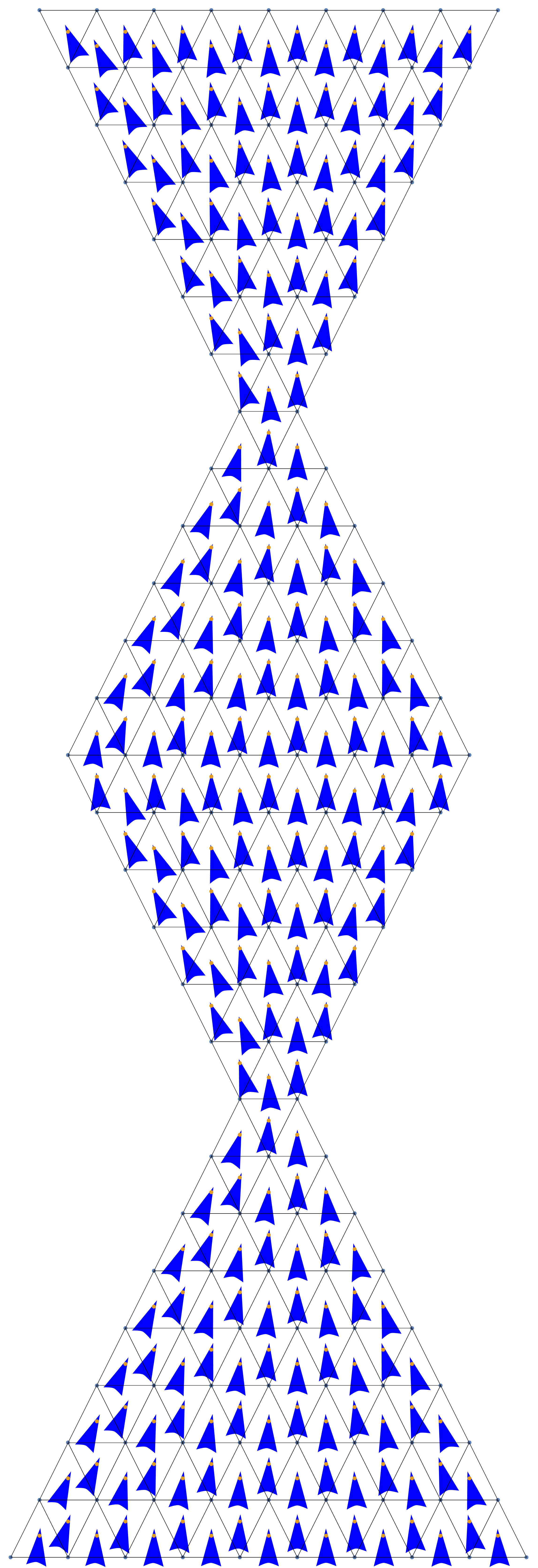}
\end{subfigure}
\caption{A smooth (left) and discrete (three copies on the right) representation of a ``conical torus" with the three discrete vector fields of $\tilde{S}$ corresponding to the three lowest eigenvalues (increasing from left to right).}
\label{fig:cone-torus}
\end{figure}

\begin{figure}
\centering
\includegraphics[width=0.7\linewidth]{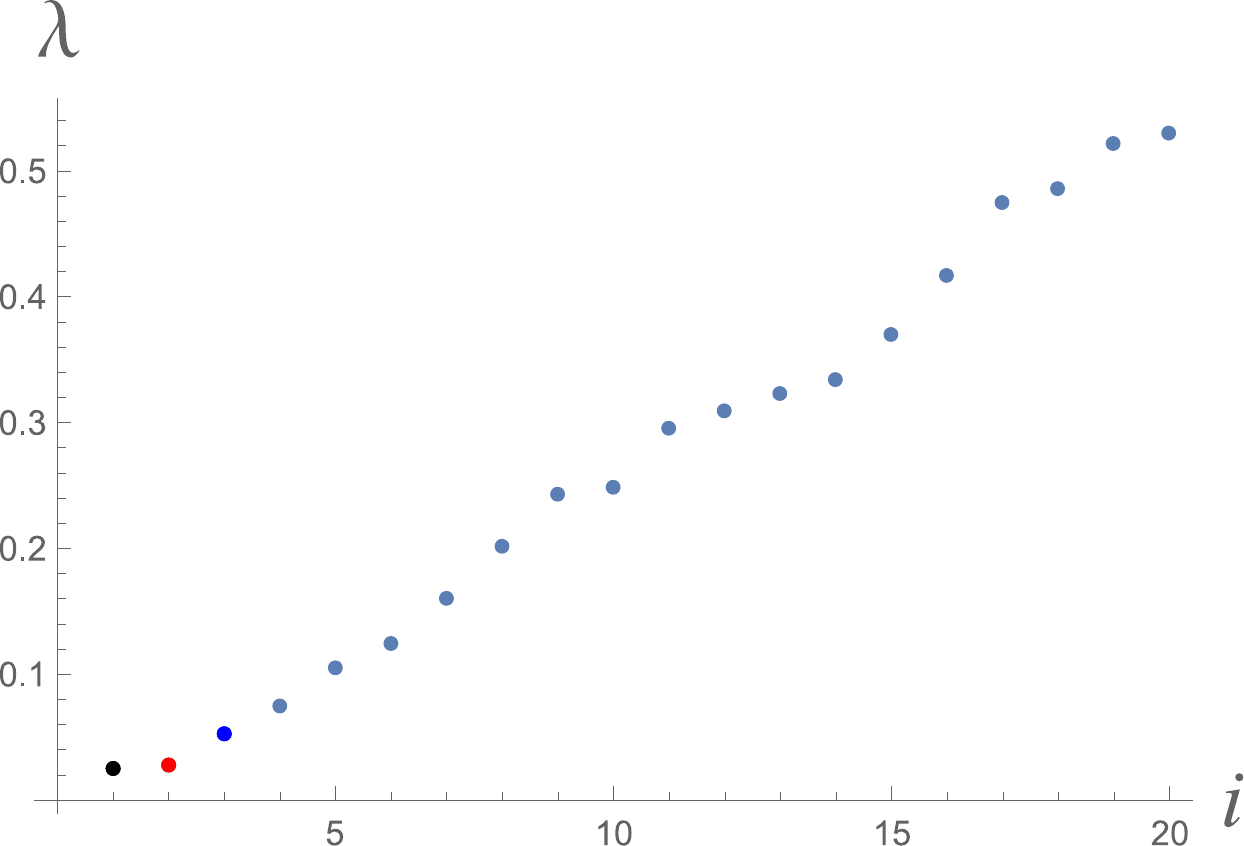}
\caption{The twenty lowest eigenvalues $\lambda_i$ (labelled on the horizontal axis) of the spectrum of the Killing energy $E(\omega)$ for a discrete conical torus. The colours of the lowest three eigenvalues correspond to the colours of the discrete vector fields in \fref{fig:cone-torus}.}
\label{fig:cone-torus-spectrum}
\end{figure}

Another interesting observation about the DAKVFs can be made by studying conical geometries of toroidal topology, like the one illustrated in \fref{fig:cone-torus}. The leftmost figure is such a ``conical torus'', a geometry whose metric is given by eq.\ \eqref{eq:cone-torus}. The top and bottom are periodically identified. The three diagrams on the right are copies of a discrete triangulation which emulates the smooth (up to the smallest and largest rings) geometry. Each shows a discrete vector field corresponding to one of the three lowest eigenvalues of $\tilde{S}$ (see \fref{fig:cone-torus-spectrum}). In the illustrations of the discrete conical torus, the top and bottom and the left and right sides are identified periodically. The conical torus admits one exact Killing vector field $\xi$.

\begin{figure}
\centering
\includegraphics[width=0.6\linewidth]{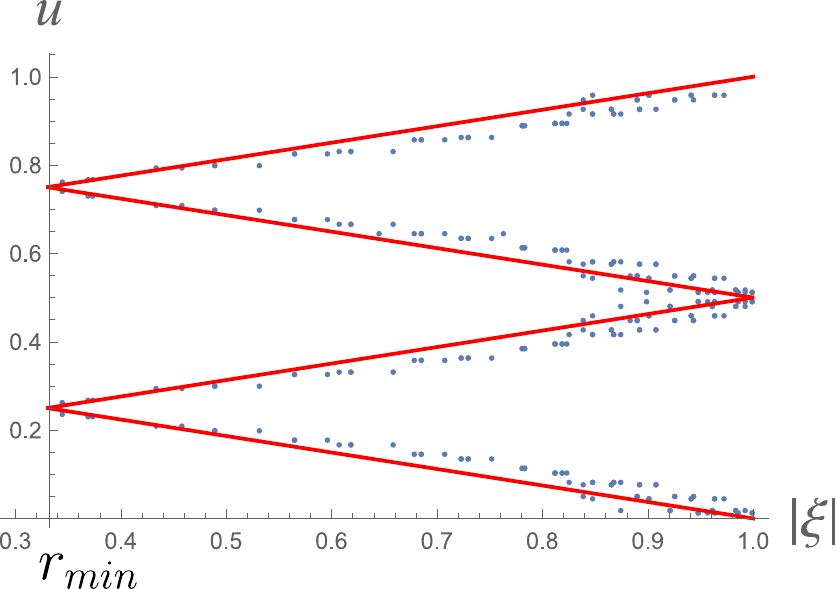}
\caption{A comparison of the continuum norm $|\xi|$ (red line) of the Killing vector field $\xi$ of the metric given by eq. \eqref{eq:cone-torus}, and the norm of the leftmost discrete vector field in \fref{fig:cone-torus}, which is the discrete analogue of $\xi$.}
\label{fig:cone-torus-norm}
\end{figure}

The continuum metric of the conical torus in the coordinates $x=(u,v)$, $0\leq u \leq 1$, $0\leq v \leq 1$ with $u(0)=u(1)$ and $v(0)=v(1)$ can be written as
\begin{equation}
g_{\mu\nu}=\left(\begin{matrix}
R(u)^2 + 2 \partial_u R(u)^2 & 0\\
0 & R(u)^2 
\end{matrix}\right),
\label{eq:cone-torus}
\end{equation}
where
\begin{equation}
R(u)= 
\begin{cases}
1 - 4(1-r_{min})u,& \ 0\leq u < \frac{1}{4}\\
r_{min}+(1-r_{min})(4u-1),& \ \frac{1}{4}\leq u < \frac{1}{2}\\
1 - (1-r_{min})(4u-2),& \ \frac{1}{2}\leq u < \frac{3}{4}\\
r_{min}+(1-r_{min})(4u-3),& \ \frac{3}{4}\leq u < 1
\end{cases}. \notag
\end{equation}
In this expression, $r_{min}$ is the minimum size of the geometry in the $v$-direction (horizontal in \fref{fig:cone-torus}), which is the length of the circle where the conical geometry is pinched most strongly. The $u$-direction is the vertical direction in \fref{fig:cone-torus}. The conical torus admits one exact Killing vector field $\xi\! =\!(0,1)$, with norm $|\xi|=R(u)$. \fref{fig:cone-torus-norm} shows a comparison of the norm of the continuum Killing vector field and of the DAKVF of the analogous discrete geometry.

We observe that the norm of the discrete vector field closely follows that of its continuum counterpart. In \fref{fig:cone-torus}, we see that the DAKVF with lowest Killing energy (the blue horizontally oriented vector field) of the discrete conical torus generates a horizontal flow comparable to $\xi$. We also observe that for the case of the discrete conical torus, where one symmetry is broken more strongly (the $u$-direction for the conical torus), the second-lowest DAKVF shows topologically different properties than what is expected for an exact Killing vector field on the torus.

Exact Killing vector fields $\xi$ on $T^2$ are harmonic ($\Delta\xi\! =\! 0$ in eq.\ \eqref{eq:hodge-decomposition}), i.e. they are vortex-free. \fref{fig:cone-torus} illustrates that the DAKVF with the second-lowest Killing energy (the red discrete vector field) is not vortex-free, because the vector field changes direction along the vertical axis. However, the corresponding Killing energy is not significantly higher. The Killing energy of the vortex-free vector field is pushed to a higher position in the spectrum of the Killing energy.

In \fref{fig:cone-torus}, this is the rightmost discrete vector field (in blue), which corresponds to the third-lowest eigenvalue. We conclude that the eigenvectors of $\tilde{S}$ corresponding to the lowest eigenvalues do not necessarily have the same topological properties as expected from exact Killing vectors if the symmetries are broken sufficiently strongly.

In the context of two-dimensional triangulations of the torus, we have observed this behaviour for many other geometries that are far from admitting any exact Killing vector fields. We have systematically studied discrete geometries of toroidal topology by performing a sequence of Pachner moves and following the behaviour of the lowest eigenvectors as a function of the moves. On these geometries we saw a similar behaviour to that on the conical torus.

We have made several attempts to define a robust measure for the relation between the low-lying spectrum of the Killing energy $\tilde{S}$ and approximate Killing symmetries of the two-torus, away from the realm of small perturbations. For example, we have studied whether gaps in the spectrum, in the sense of relative differences between the first few eigenvalues, are a robust measure of approximate symmetries. Other attempts were based on the discrete flow and the vorticity of the vector fields. Due to the complex relation between low-lying eigenvalues and eigenvectors of $\tilde{S}$ and the properties of a general geometry it is difficult to find a well-motivated definition of symmetry-breaking in terms of the spectrum and eigenvectors of $\tilde{S}$.

A more systematic study of the role of vortex-free vector fields in the spectrum of $\tilde{S}$ and their relation to approximate Killing vectors is beyond the scope of this work. Instead, in what follows we will propose a specific observable for quantum gravity, based on the eigenvectors of $\tilde{S}$, motivated by the results of this section. Although we presently do not have a general understanding of the complex interplay between the eigenvectors of $\tilde{S}$ and the properties of a general geometry, such an observable can potentially provide new insights into the role of (approximate) symmetries in theories of quantum gravity. We will present an analysis of a specific choice of such an observable in \sref{sec:symmetries-qg}.

\section{Symmetries in two-dimensional \texorpdfstring{quantum \\gravity}{quantum gravity} on a torus}
\label{sec:symmetries-qg}

\subsection{A new observable}

We now return to the question that motivated this study. An important question in any approach to quantum gravity is whether we can define observables, which can capture properties of the theory that, in an appropriate limit, are related to continuum notions such as (approximate) Killing symmetries. In \sref{sec:example-geometries} we concluded that the numerical value of the lowest eigenvalues in the spectrum of the Killing energy $\tilde{S}$ most likely is not a sufficiently robust tool to analyse approximate symmetries when the geometries under consideration are far from regular. However, the associated vector fields could potentially have more suitable properties. We have investigated whether a specific observable based on the eigenvectors of the discrete Killing energy $\tilde{S}$ may be more useful as a tool to analyse approximate symmetries.

First we need to define what is meant by the vorticity of a vector field (dual to) $\omega$. For a differentiable closed manifold $\mathcal{M}$ and a vector field $\omega$ with a collection of isolated zeroes $\omega(p_i)=0$ at points $\{ p_i \}$, the index $\mathrm{Ind}(p_i)$ of an isolated zero $\omega(p_i)$ is defined as follows. In terms of a local coordinate patch around $p_i$, $\mathrm{Ind}(p_i)$ is the degree of the map $\hat{n}: T_x \mathcal{M}\rightarrow S^{n-1}$ for all $x \in \partial B(p_i)$, where $B(p_i)$ is a closed (topological) ball around $p_i$ that does not include any other zeroes of $\omega$. Furthermore, $\partial B(p_i)$ is the boundary of the ball $B(p_i)$, $T_x \mathcal{M}$ is the tangent space of $\mathcal{M}$ at the point $x$, and $S^{n-1}$ is the $(n-1)$-sphere. Explicitly, the map is given by $\hat{n}(\omega(x))=\omega(x)/\left|\omega (x)\right|$, where we define the norm $|\alpha|=\sqrt{\alpha_\mu\alpha^\mu}$ for a general vector field $\alpha^\mu$. In more pedestrian terms, the index of an isolated zero is the winding number of the map from the vector field evaluated on an arbitrary $(n-1)$-dimensional shell $\partial B(p_i)$, which only encloses the zero at $p_i$, to the $(n-1)$-sphere $S^{n-1}$. The vorticity of a vector field $\omega$ is given by the collection of the isolated zeroes $p_i$ and the indices $\mathrm{Ind}(p_i)$. The Poincar\'e-Hopf theorem states that
\begin{equation}
\sum_i \mathrm{Ind}(\omega(p_i)) = \chi(\mathcal{M}),
\label{eq:poincare-hopf}
\end{equation}
where the sum runs over the isolated zeroes $p_i$ of $\omega$ and $\chi(\mathcal{M})$ is the Euler characteristic of $\mathcal{M}$. The vorticity of a discrete vector field is defined in analogy with the Poincar\'e-Hopf theorem, where it should be noted that a discrete analogue of eq.\ \eqref{eq:poincare-hopf} is not sensitive to variations of a vector field $\omega$ on scales smaller than the lattice spacing.

In \sref{sec:example-geometries}, we discussed an interesting property of the eigenvectors of the discrete Killing operator of a specific simplicial geometry, the conical torus. The exact Killing vector fields on a flat torus have zero vorticity. The discrete vector fields $\omega$ with zero vorticity are present in the set of eigenvectors of $\bar{S}$, but correspond to eigenvalues higher than the lowest two eigenvalues. One way of analysing the vorticity of a vector field is with the help of a Hodge decomposition, because zeroes of a vector field $\omega$ correspond to extrema of the potentials $\phi$ and $\psi$ in the decomposition in eq.\ \eqref{eq:hodge-decomposition}. However, we have seen that the discrete analogue of the Hodge decomposition of eq.\ \eqref{eq:hodge-decomposition} seems to be too sensitive to the details of the discretisation to be useful for our purposes. We therefore propose a different observable to capture the vorticity of the discrete vector fields.

For a continuum manifold $\mathcal{M}$ which admits both a harmonic form $(\omega_h)_\mu$ and a one-form $(\omega_\xi)_\mu$ dual to a Killing vector field, we know that their inner product is constant \cite{On harmonic and Killing vector fields}
\begin{equation}
(\omega_h(x))_\mu(\omega_\xi(x))^\mu=c, \quad x\in\mathcal{M}, \quad c \in \mathbb{R}.
\label{eq:xi-h}
\end{equation}
On a two-torus we can make an even stronger statement. Given a two-dimensional closed manifold $\mathcal{M}$ that admits a nowhere vanishing Killing vector $\xi^\mu$, we can always make a coordinate choice $x=(x_1,x_2)$ in which $\xi$ is a basis vector\footnote{A coordinate chart can be based on a Killing vector field that vanishes at certain points, but then the chart will not cover these points.},
\begin{equation}
\xi^\mu = \partial_{x_2}^\mu,
\label{Kbasis}
\end{equation}
and the metric on $\mathcal{M}$ takes the form
\begin{equation}
g_{\mu\nu}=\left( \begin{matrix}
f_{11}(x_1)&f_{12}(x_1)\\
f_{12}(x_1)&f_{22}(x_1)
\end{matrix} \right).
\label{Kbasismetric}
\end{equation}
The divergence of $\xi^\mu$ vanishes because $\xi^\mu$ is a Killing vector. The exterior derivative of the dual $\xi_\mu$ is equal to
\begin{equation}
(d\xi)_{\mu \nu} = \left( \begin{matrix}
0&\partial_{x_1}f_{22}\\
-\partial_{x_1}f_{22}&0
\end{matrix} \right).
\label{dxi}
\end{equation}
As long as $f_{22} \neq 0$ everywhere on $\mathcal{M}$, we can consider $(\omega_h)_\mu=\xi_\mu/f_{22}$ and notice that $(d \omega_h)_{\mu \nu}=0$. Calculating $\nabla^\mu(\omega_h)_\mu$, we also see that the divergence of $(\omega_h)_\mu$ vanishes. We conclude that $(\omega_h)_\mu$ is a harmonic one-form. The one-form $\xi_\mu(x)=f_{22}(x) (\omega_h)_\mu(x)$ is therefore equal to a harmonic one-form $(\omega_h)_\mu$, up to an $x$-dependent rescaling by $f_{22}^{-1}(x)$, if $f_{22}(x) \neq 0$ and a coordinate system exists that is well-defined globally. This is true when $\mathcal{M}$ has toroidal topology. To exploit this property, we will turn to the special case of geometries of toroidal topology. In the remainder of this section we will also call two one-forms that only differ by a local rescaling ``parallel''. We have therefore shown that the one-form $\xi_\mu$ dual to a Killing vector field $\xi^\mu$ on a two-torus is always parallel to a harmonic form.

On a two-torus $\mathcal{M}$, every locally normalised one-form $\xi_\mu/|\xi|$ constructed from a Killing vector field $\xi^\mu$ is equal to a locally normalised one-form $(\omega_h)_\mu/|\omega_h|$ constructed from a harmonic one-form $(\omega_h)_\mu$, where we define the norm $|\alpha|=\sqrt{\alpha_\mu\alpha^\mu}$ for a general one-form $\alpha_\mu$. From Hodge theory, we know that the number of linearly independent harmonic forms of a manifold is a topological invariant. The number of harmonic $k$-forms is equal to the $k$th Betti number. More specifically, this means that the number of harmonic one-forms is equal to the rank of the first homology group. On the torus, there always exist two linearly independent harmonic forms $(\omega^1_h)_\mu$ and $(\omega^2_h)_\mu$, because the rank of the first homology group is equal to two. Given a Killing vector field on a torus, we can therefore always find a harmonic one-form for which the locally normalised Killing vector field and the locally normalised vector field dual to the harmonic one-form are equal. It should be noted that although the locally normalised vector fields are equal, they are generally no longer Killing vector fields or dual to a harmonic one-form.

We propose to use the result on Killing vectors and harmonic one-forms on a two-torus to construct an observable for two-dimensional quantum gravity on the torus. We will measure the deviation from parallelism of the discrete Killing vector fields associated to the lowest eigenvalues of $\tilde{S}$ and the vector fields dual to a harmonic one-form on a discrete torus $T$.

With the two harmonic forms $(\omega^1_h)_\mu$ and $(\omega^2_h)_\mu$ on the two-torus we can define an observable $P(\omega_i)$ as a function of the eigenvector $\omega_i$ of the Killing energy $\tilde{S}$ associated to the $i$th-lowest eigenvalue. We will make use of a locally normalised linear combination $(\hat{\omega}_h)_\mu=(\omega_h)_\mu/|\omega_h|$ of the harmonic one-forms $(\omega^1_h)_\mu$ and $(\omega_h^2)_\mu$,
\begin{equation}
(\omega_h)_\mu=a(\omega^1_h)_\mu+b(\omega_h^2)_\mu,
\label{TorusHarmonic}
\end{equation}
for $a, b \in \mathbb{R}$. The expression for the proposed observable $P(\omega_i)$ is
\begin{equation}
P(\omega_i)=\max\limits_{a,b}\frac{1}{V}\int dV \ |\hat{\omega}_i, \hat{\omega}_h|,
\label{Observable}
\end{equation}
where $V$ denotes the volume of the torus. The observable is normalised to take values between zero and one, $0 \leq P(\omega_i) \leq 1$. The observable $P(\omega_i)$ is the normalised volume integral of the inner product between the locally normalised vector field $(\hat{\omega}_i)^\mu=(\omega_i)^\mu/|\omega_i|$ constructed from an eigenvector $\omega_i$ of $\tilde{S}$ and the locally normalised vector field $(\hat{\omega}_h)^\mu=(\omega_h)^\mu/|\omega_h|$ constructed from a set of two orthonormal harmonic one-forms $(\omega^1_h)_\mu$ and $(\omega^2_h)_\mu$. The integral is maximised with respect to the parameters $a$ and $b$. Note that $P(\omega_i)$ is independent of the choice of basis $\omega^1_h$ and $\omega_h^2$ for the two-dimensional space of harmonic one-forms on a torus. The observable will be set to zero at the zero points of the one-form fields $\omega_h$ and $\omega_i$, where the observable would diverge due to the normalisation. In the discrete application we have in mind, such points will not play any role.

The observable $P(\omega_i)$ is a measure of how parallel the vector field $(\hat{\omega}_i)^\mu$ is to the vector field dual to a harmonic form $(\hat{\omega}_h)^\mu$. If $(\omega_i)^\mu$ is a Killing vector field of the two-torus, the observable $P(\omega_i)$ saturates its upper bound. On a flat torus, all higher modes $\omega_i$ of $\tilde{S}$ that are not Killing saturate the lower bound. We have furthermore observed that the variation of the value of $P(\omega_i)$ under small deformations of the geometry is also small, which reflects the result derived in \cite{Beetle:2013}. It should be noted that the locally normalised one-forms $\hat{\omega}_i$ and $\hat{\omega}_h$ are in general not orthogonal with respect to the standard $L^2$-norm on forms. We therefore do not have control over the behaviour of the observable $P(\omega_i)$, beyond small perturbations of the flat torus. This implies that the interpretation of values that are neither close to zero nor one is not immediately clear. We have not been able to find a definition of an observable in terms of the standard $L^2$-norm for which there is a simple geometric interpretation on an ensemble of geometries.

Although we do not have a good understanding of the interpretation of values of $P(\omega_i)$ that are neither close to zero nor close to one, and although we have no clear interpretation of $P(\omega_i)$ for higher modes $\omega_i$ for geometries that are far from regular, we at least now have an observable that can be implemented in certain toy models of quantum gravity. The properties of $P(\omega_i)$ we discussed above have motivated us to investigate whether $P(\omega_i)$ is a suitable observable to distinguish between one-forms related to symmetries and other modes of $\tilde{S}$, when we consider an ensemble average in these toy models. The two-dimensional models we have studied are DT and CDT with toroidal topology and small perturbations around the flat torus, using a discrete analogue of $P(\omega_i)$. Note that while in the continuum the inner product $|\alpha,\beta|$ of two vector fields $\alpha^\mu$ and $\beta^\mu$ is equal to the inner product of the two corresponding one-forms $\alpha_\mu$ and $\beta_\mu$, this is not the case for the discrete one-forms $\alpha$, $\beta$ and the discrete vector fields $\alpha^\sharp$, $\beta^\sharp$ in the framework of DEC. A local inner product is only defined for the discrete vector fields $\omega^\sharp$. We will therefore define the discrete analogue of the proposed observable $P(\omega_i)$ in terms of discrete vector fields $\omega^\sharp$ instead of discrete one-form fields $\omega$.

The discrete version of \eqref{Observable} is given by
\begin{equation}
P(\omega_i)=\max\limits_{a,b}\frac{1}{N_2}\sum_{\sigma^2} \hat{\omega}_i^\sharp \cdot \hat{\omega}_h^\sharp.
\label{DiscreteObservable}
\end{equation}
In the discrete observable $P(\omega_i)$, the discrete vector field $\hat{\omega}_i^\sharp$ is the locally (per 2-simplex $\sigma^2$) normalised vector field,
\begin{equation}
\hat{\omega}_i^\sharp(\sigma^2)\equiv\frac{\omega_i^\sharp(\sigma^2)}{|\omega_i^\sharp(\sigma^2)|}
\label{DiscreteObservable1}
\end{equation}
constructed from the discrete one-form $\omega_i$ associated to the $i$th-lowest eigenvalue of $\tilde{S}$. The discrete vector field $\hat{\omega}_h^\sharp$ is the locally (per 2-simplex $\sigma^2$) normalised vector field,
\begin{equation}
\hat{\omega}_h^\sharp(\sigma^2)\equiv\frac{\omega_h^\sharp(\sigma^2)}{|\omega_h^\sharp(\sigma^2)|},
\label{DiscreteObservable2}
\end{equation}
constructed from the harmonic discrete one-form $\omega_h=a\omega_h^1+b\omega_h^2$ for which $\omega_h^1$ and $\omega_h^2$ solve the discrete Laplace equation
\begin{equation}
\Delta\omega^1_h=0, \quad \Delta\omega^2_h=0, \quad \Delta=(d\delta+\delta d).
\end{equation}
The norm and inner product for the discrete expressions in eqs.\ \eqref{DiscreteObservable}, \eqref{DiscreteObservable1}, and \eqref{DiscreteObservable2} are defined with respect to the flat metric on the 2-simplex $\sigma^2$. The sum in the discrete expression for $P(\omega_i)$ runs over all 2-simplices $\sigma^2$ in the discrete geometry $T$. This observable is quite costly to calculate for a discrete geometry, due to the maximisation with respect to $a$ and $b$. To simplify the computations we will define a related observable. We choose two discrete harmonic forms $\omega^1_h$ and $\omega^2_h$ at random from the two-dimensional vector space on a torus spanned by $\omega^1_h$ and $\omega^2_h$. We then compute the standard deviations $\sigma_T(\hat{\omega}_i^\sharp \cdot \hat{\omega}_h^{1\sharp})$ and $\sigma_T(\hat{\omega}_i^\sharp \cdot \hat{\omega}_h^{2\sharp})$ of the distribution  over 2-simplices $\sigma^2 \in T$ of the inner products $\hat{\omega}_i^\sharp(\sigma^2) \cdot \hat{\omega}_h^{1\sharp}(\sigma^2)$ and $\hat{\omega}_i^\sharp(\sigma^2) \cdot \hat{\omega}_h^{2\sharp}(\sigma^2)$ per 2-simplex. The observable $\tilde{P}(\omega_i)$ is defined as the larger one of the two standard deviations,
\begin{equation}
\tilde{P}(\omega_i)=\max \lbrace \sigma_T( \hat{\omega}_i^\sharp \cdot \hat{\omega}_h^{1\sharp}),\sigma_T( \hat{\omega}_i^\sharp \cdot \hat{\omega}_h^{2\sharp}) \rbrace.
\label{DiscreteObservable3}
\end{equation}
The standard deviation of the inner product is also a good measure for how parallel the vector fields $\omega^\sharp_i$ and $\omega^\sharp_h$ are. For the flat torus the observable $\tilde{P}(\omega_i)$ is equal to zero for the two discrete Killing vector fields and close to one for all other eigenvectors of $\tilde{S}$. The observables $P(\omega_i)$ and $\tilde{P}(\omega_i)$ are strongly correlated for random simplicial geometries of toroidal topology, in the numerical range where we could compute both.

\subsection{Results and discussion}
\label{ssec:results-discussion}
In this section we discuss the results of a measurement of the expectation value of $\tilde{P}(\omega_i)$ with the numerical methods described in Sec.\ \ref{sec:motivation}. We have sampled the respective ensembles for three types of two-dimensional toroidal geometries with Monte Carlo methods weighted by the exponentiated Regge action $e^{-S_R[T]}$ evaluated on the triangulation $T\in \mathcal{T}$. We will compare the proposed observable between CDT with $\mathcal{T}=\mathcal{T}_{CDT}$, DT with $\mathcal{T}=\mathcal{T}_{DT}$ and small perturbations on a flat torus. The small perturbations are a subset of the configuration space $\mathcal{T}_{CDT}$ of CDT geometries, where no vertex order is smaller than five or larger than six. The small perturbations are obtained by performing $\sqrt{N_2/2}$ volume preserving Pachner moves at random points on the flat torus.

\begin{figure}[t!]
\centering
\begin{tabular}[c]{ccc}
\centering
\begin{subfigure}[t]{0.3\textwidth}
\includegraphics[width=\linewidth]{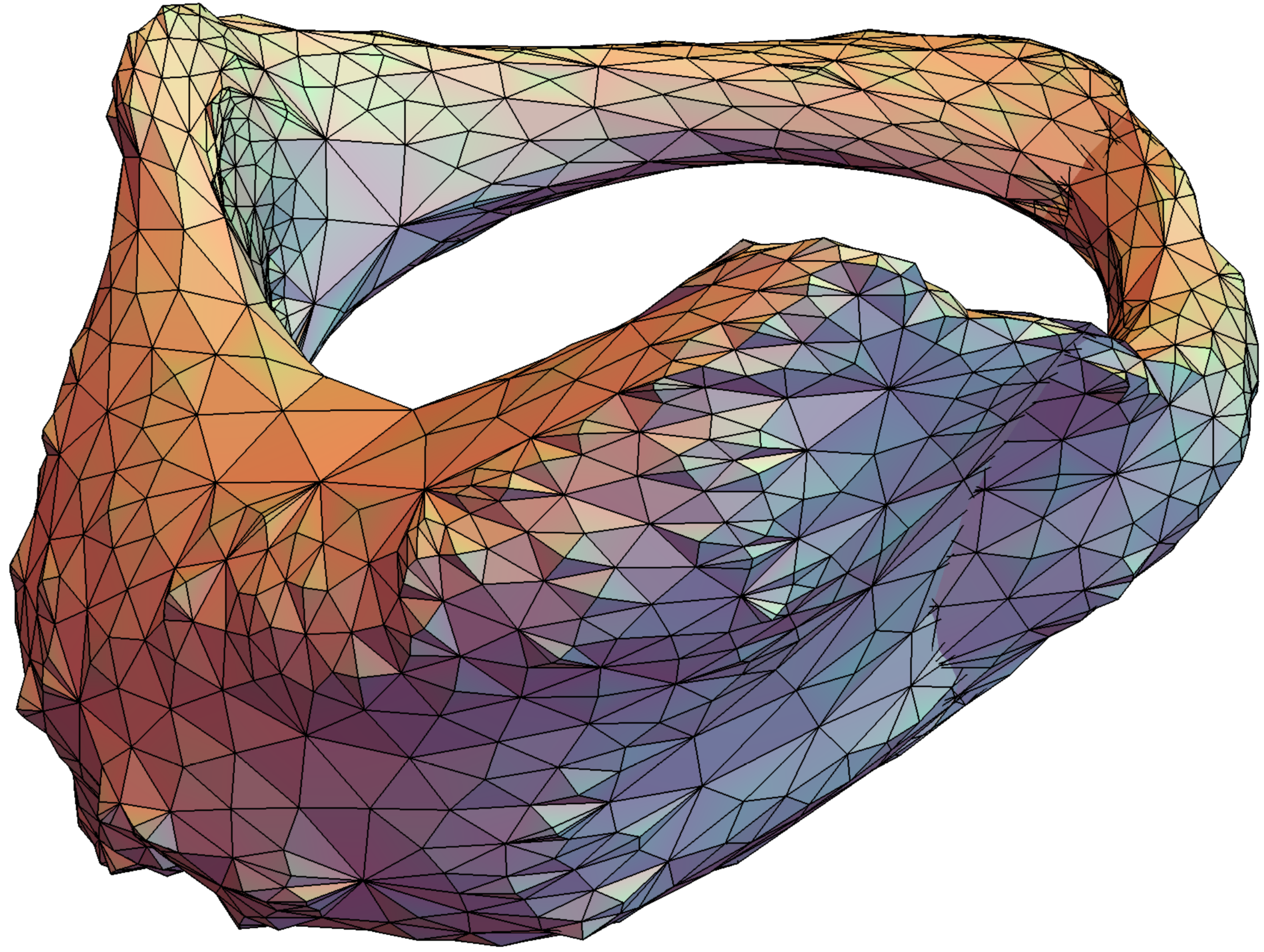}
\caption{CDT}
\label{fig:CDT}
\end{subfigure}&
\begin{subfigure}[t]{0.3\textwidth}
\includegraphics[width=\linewidth]{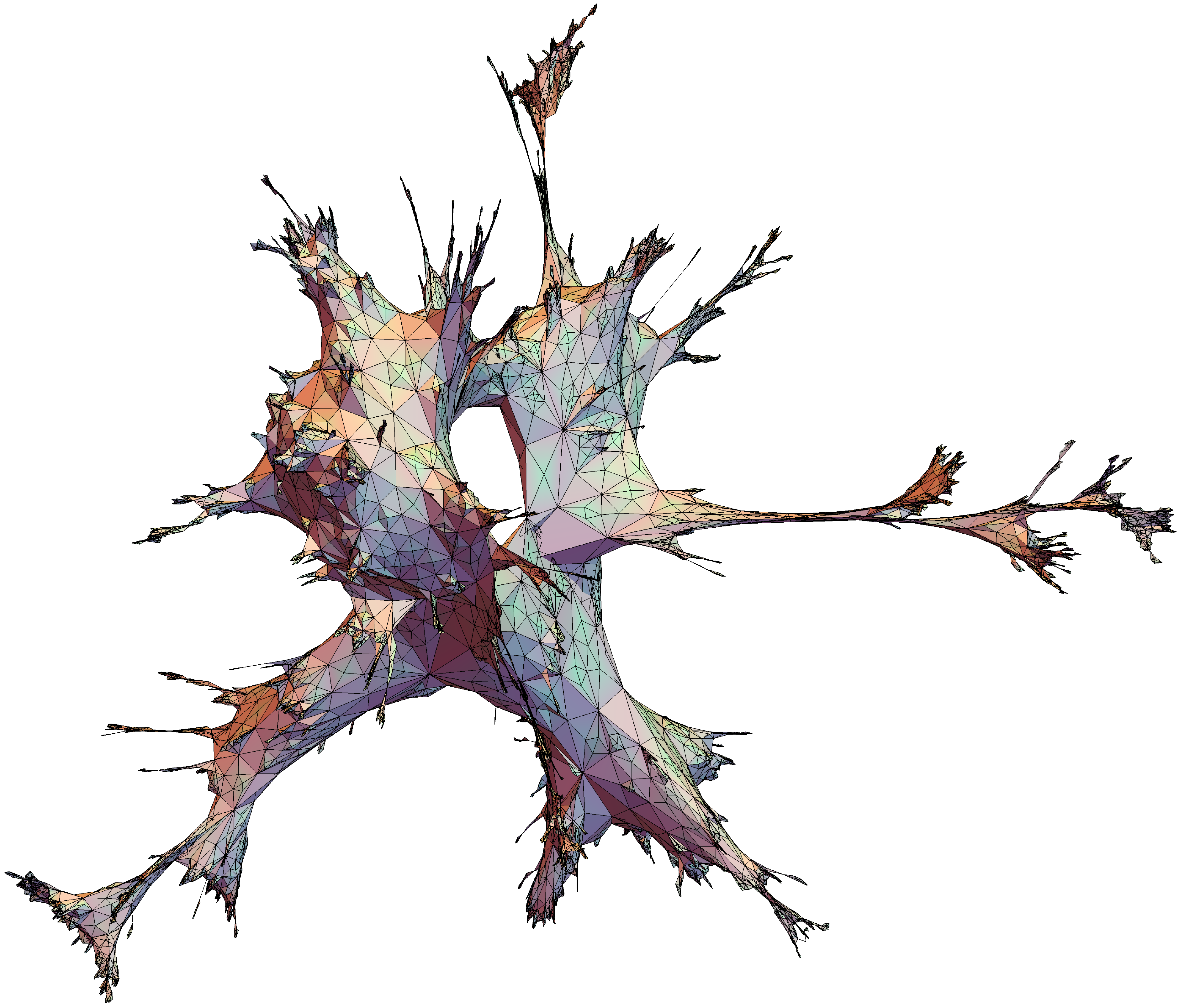}
\caption{DT}
\label{fig:DT}
\end{subfigure}&
\begin{subfigure}[t]{0.3\textwidth}
\includegraphics[width=\linewidth]{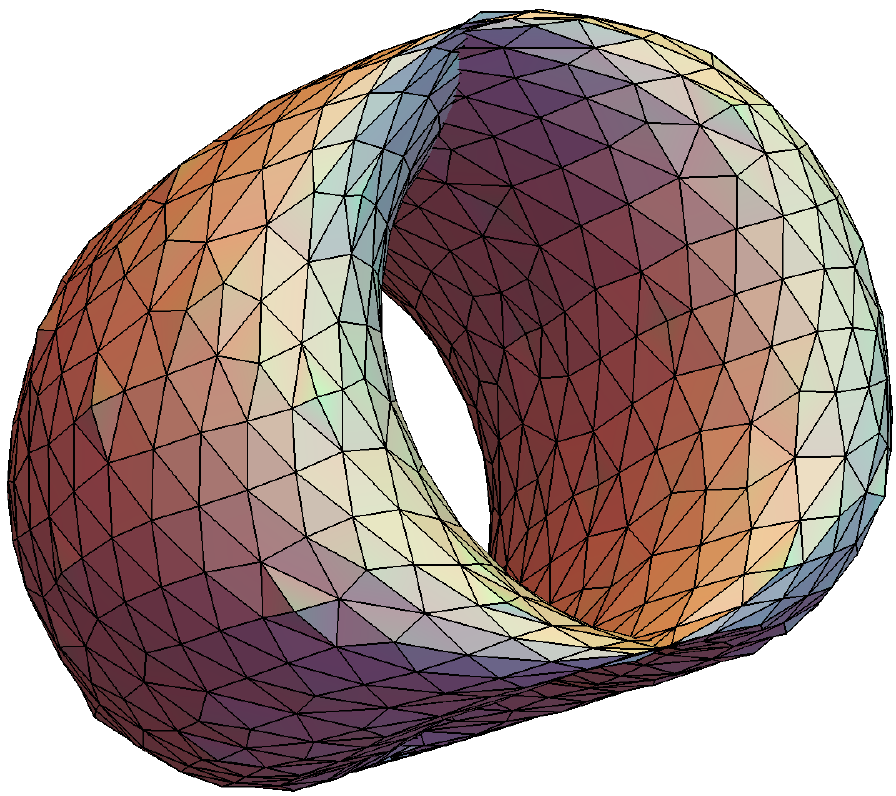}
\caption{Perturbations}
\label{fig:Per}
\end{subfigure}
\end{tabular}
\caption{Typical geometries with toroidal topology from ensembles of geometries for CDT, DT and small perturbations on flat space for $N_2=800$. The geometries are coloured with a gradient from green to red for vertices of low to high vertex order. The apparent lengths of the links are an artefact of the embedding. In reality, all simplices are equilateral.}
\end{figure}

When we construct the DAKVFs for a typical DT geometry we encounter a complication. In contrast to the continuum, the definition of the DAKVF is not necessarily positive definite for simplicial complexes that include vertices with vertex order smaller than four. Such geometries do not appear in CDT and for small perturbations on a flat torus, but vertices of order three can appear in DT geometries. To avoid this issue, we adopt a restricted version of DT for which we do not allow triangulations with vertex order smaller than four. Typical geometries in each of these ensembles are depicted in Fig. \ref{fig:Per}.

\begin{figure}
\centering
\begin{tabular}[c]{cc}
\begin{subfigure}[b]{0.47\textwidth}
\includegraphics[width=\linewidth]{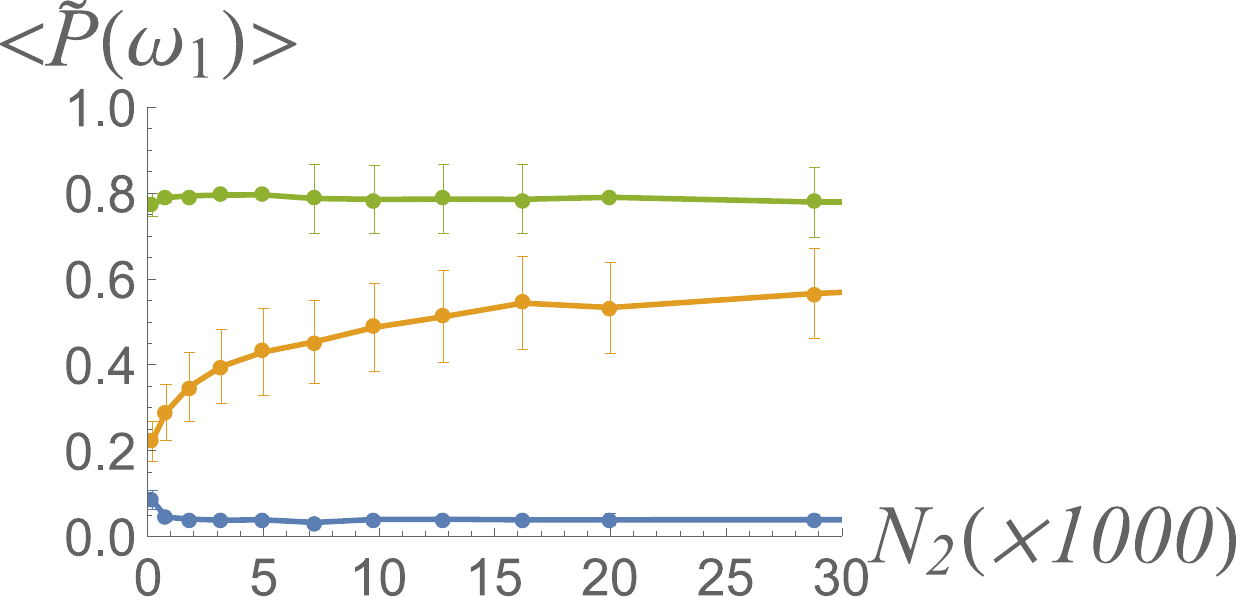}
\end{subfigure}&
\begin{subfigure}[t]{0.47\textwidth}
\includegraphics[width=\linewidth]{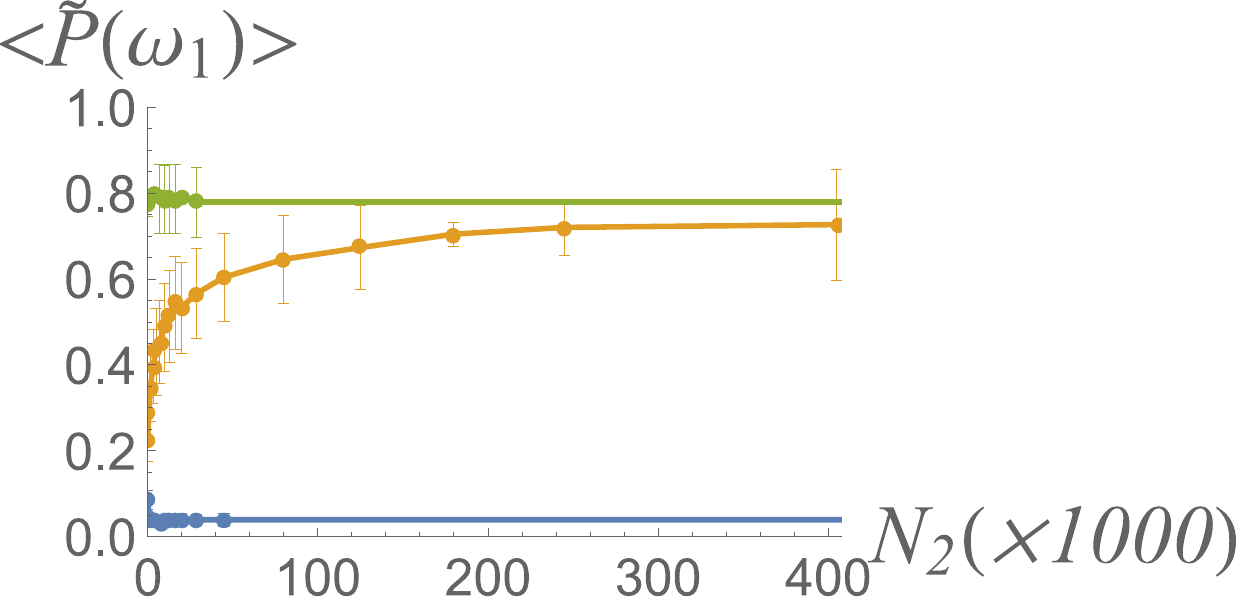}
\end{subfigure}\\
\begin{subfigure}[b]{0.47\textwidth}
\includegraphics[width=\linewidth]{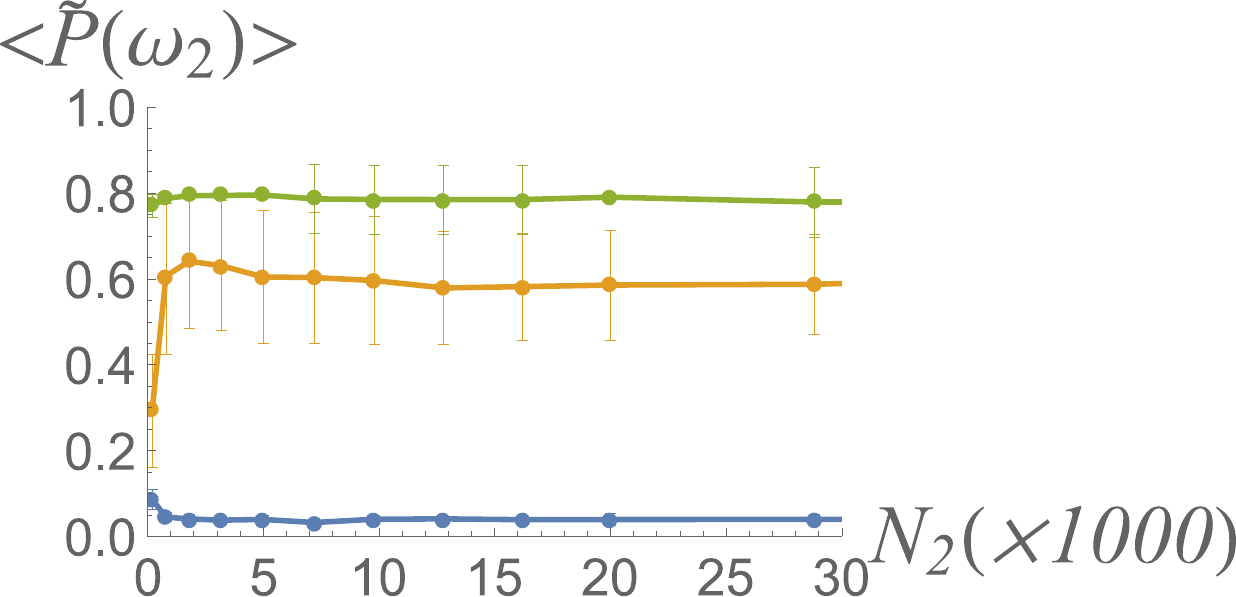}
\end{subfigure}&
\begin{subfigure}[t]{0.47\textwidth}
\includegraphics[width=\linewidth]{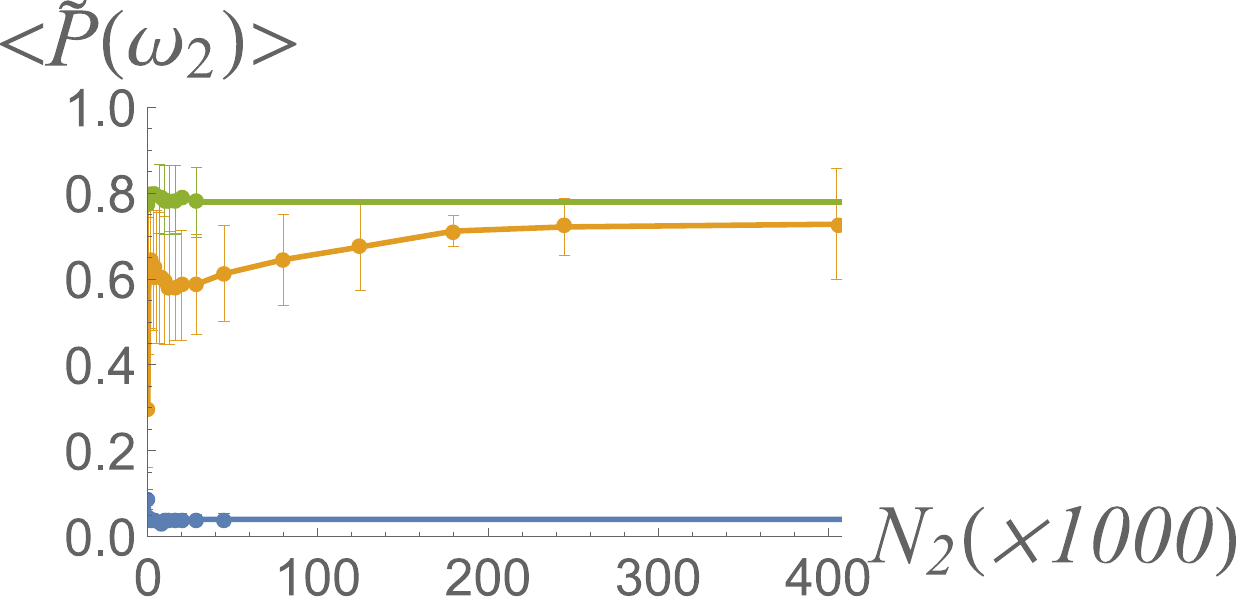}
\end{subfigure}\\
\begin{subfigure}[b]{0.47\textwidth}
\includegraphics[width=\linewidth]{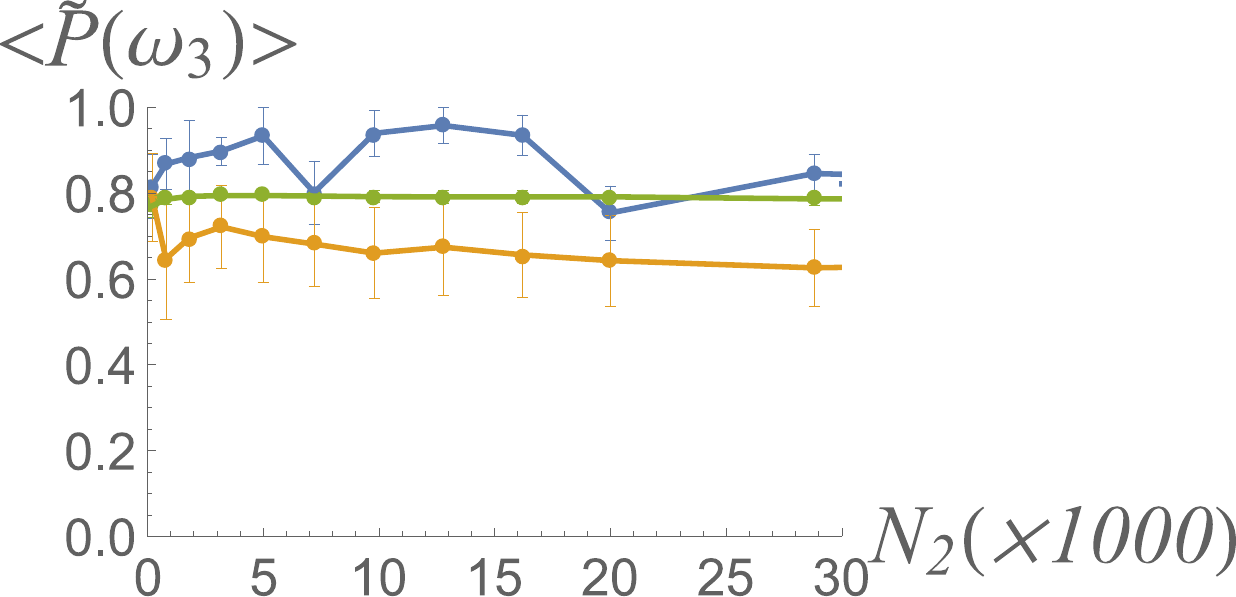}
\end{subfigure}&
\begin{subfigure}[t]{0.47\textwidth}
\includegraphics[width=\linewidth]{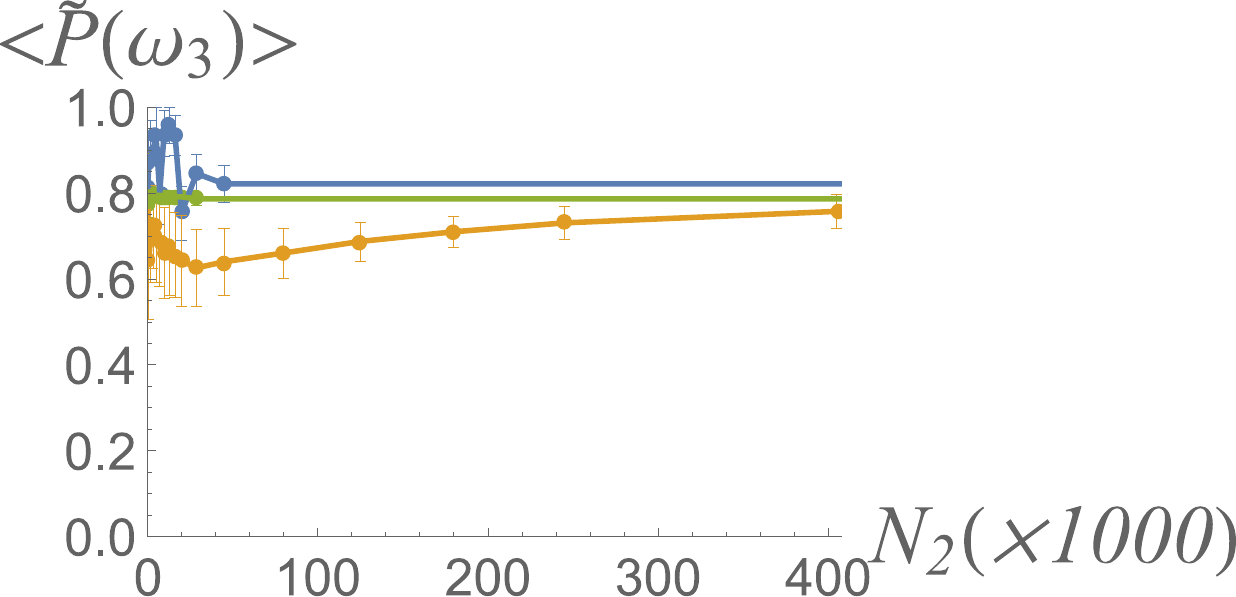}
\end{subfigure}
\end{tabular}
\caption{Measurements of the expectation value $\langle \tilde{P}(\omega_i) \rangle$ as a function of the number of triangles $N_2$ for CDT (orange), DT (green), and small perturbations (blue). The right column shows an extended range for the scaling of CDT, assuming that the values of DT and the small perturbations remain approximately constant.}
\label{fig:scaling-data}
\end{figure}

The average of $\tilde{P}(\omega_i)$ over a small sample of the ensemble at fixed number of triangles $N_2$ approximates the expectation value $\langle\tilde{P}(\omega_i)\rangle_{N_2}$. We have repeated the measurements for different values of $N_2$. We can therefore analyse the scaling of the expectation value as a function of $N_2$ and attempt to extract the infinite-volume limit. For CDT and the small perturbations, the geometries consist of $\sqrt{N_2/2}$ time slices. The results are summarised in \fref{fig:scaling-data}, combining the data for CDT (orange), DT (green) and the small perturbations (blue). The figure shows the measured values $\langle \tilde{P}(\omega_i)\rangle_{N_2}$ for various total number of triangles $N_2$, of the three eigenvectors $\omega_1$, $\omega_2$ and $\omega_3$ of the Killing energy operator $\tilde{S}$, corresponding to the lowest three eigenvalues $\lambda_1<\lambda_2<\lambda_3$. Each data point is an average over about $100$ geometries with the corresponding standard deviation. The lines that connect the data points are merely added to guide the eye. The plots in the right column show measurements for an extended range for the scaling of CDT, assuming that the values of DT and the small perturbations do not change in this range.

The expectation value of $\tilde{P}(\omega_i)$ for DT is consistently at around $0.8$ for all three eigenvectors $\omega_1$, $\omega_2$ and $\omega_3$. We conjecture that a value close to one is a sign that there are no approximate symmetries present. For the small perturbations, the expectation value $\langle\tilde{P}(\omega_i)\rangle_{N_2}$ is consistently close to zero for the eigenvectors $\omega_1$ and $\omega_2$ and close to one for $\omega_3$. This signals that there are two DAKVFs present. The behaviour of the observable for CDT is more complicated. For small geometries, it seems that the expectation value for the eigenvector corresponding to the lowest values $\langle\tilde{P}(\omega_1)\rangle_{N_2}$ is close to zero, while the expectation values for $\omega_2$ and $\omega_3$ are significantly higher. The lower value of  $\langle\tilde{P}(\omega_1)\rangle_{N_2}$ could indicate that the breaking of one of the two exact symmetries of the torus is less severe in CDT. We might conclude that one approximate Killing vector field is present. Explicit investigations of the corresponding discrete vector fields $\omega_1^\sharp$ indicate that this is related to the layered structure of CDT.

However, when extrapolated to infinite volume, the expectation values for CDT tend to those of DT. The anisotropy implied by the presence of one DAKVF in CDT seems to be a finite-size effect. This could be seen as a desirable result, indicating that the layered structure of CDT does not leave an explicit physical imprint on the geometry in the continuum limit.

We conclude that investigations of the new observable $\tilde{P}(\omega_i)$ did not find any discrete approximate Killing vector fields in two toy models of quantum gravity, namely, DT and CDT on a torus. Since quantum fluctuations are dominant in these theories, this lies within our expectations. For the case of small perturbations on a flat torus, we do find two discrete approximate Killing vectors. At small volumes we also saw some evidence of a discrete approximate Killing vector field in CDT. Although this feature does not persist at larger volumes, it suggests that it may be possible to study direction-dependent properties in a quantum theory of gravity with the observable $\tilde{P}(\omega_i)$. One difficulty in understanding this new observable is the interpretation of values of $\langle\tilde{P}(\omega_i)\rangle$ that are neither close to zero nor close to one. Without a scale of comparison, these values do not have any immediate interpretation in terms of the symmetries of the underlying quantum geometries.

The introduction of the observable $\tilde{P}(\omega_i)$ is an attempt to find a quantitative measure for the presence of approximate isometries on a geometry. The construction based on the lowest eigenvectors of the Killing energy operator $\tilde{S}$ only seems to be useful in the context of small perturbations. For a definition that is more widely applicable, it is likely that a more thorough understanding of the spectrum of $\tilde{S}$ is necessary. One idea could be to examine how the exact Killing vector fields ``move" in the spectrum under continuous deformations of the geometry. The Killing energy associated to such a vector field might be a better measure of the presence of approximate isometries.

However, a full analysis of the spectrum of differential operators like $\tilde{S}$ is difficult and also may not be necessarily relevant in the quantum theory, where one is interested in the behaviour of averages or expectation values in a continuum limit. The idea of reconstructing properties of a geometry from the spectrum of a differential operator has a long history \cite{Kac:1966} and generally goes by the name of spectral geometry. The Laplace-Beltrami operator has been extensively studied with this goal in mind, see for example \cite{Reuter:2005}. In \cite{Kempf:2012}, an algorithm was presented to reconstruct a two-dimensional geometry with spherical topology from the spectrum of the scalar Laplace-Beltrami operator. It was also argued in \cite{Kempf:2012} that higher-order differential operators are necessary to achieve a similar result in higher dimensions. Let us also point out that the low-lying spectrum and eigenvectors of the scalar Laplacian have recently been studied as an order parameter in the context of CDT \cite{Clemente:2018,Clemente:2019}. Another well-known related framework is non-commutative geometry \cite{Connes:2017}, where a generalised notion of geometry is encoded in the spectrum of a Dirac-type operator.

The Killing energy operator $S$ is closely related to the link Laplace-Beltrami operator. The first two terms in $S$ in eq.\ \eqref{eq:killing-operator} contain the link Laplacian. One important difference between the link Laplacian and $S$ is that the lowest modes of the link Laplacian only contain topological information determined by the first Betti number, while the lowest modes of $\tilde{S}$ also carry metric information.

A possible next step is to construct the discrete Killing energy in geometries of higher dimension. To achieve this, a notion of a discrete Ricci tensor must be defined. A Ricci tensor based on links, like the one constructed in \cite{Alsing:2011}, seems most relevant for an operator on discrete one-forms, but further investigations are necessary. One of our conclusions was that the interpretation of the DAKVFs as approximate symmetries only seems warranted when the fluctuations are small. This suggests that a study of symmetry-reduced models could be promising, such as those discussed in \cite{Dittrich:2006,Dittrich:2002}. A possible application in a non-perturbative context is phase $C$ of CDT in four dimensions with $T^4$-topology. The spatial slices are locally very irregular, but the variations of the volume profile are relatively small. It could be that this approximately regular structure is visible in terms of the new observable $P(\omega_i)$ and perhaps is reflected in the presence of an approximate Killing vector in the time direction.

\section*{Acknowledgements}
We want to thank Timothy Budd for helpful discussions on Discrete Exterior Calculus. We also want to thank Renate Loll for her advice throughout the development of the ideas in this article. M.R.\ was supported in part by the Perimeter Institute for Theoretical Physics. Research at Perimeter Institute is supported by the Government of Canada through the Department of Innovation, Science, and Economic Development, and by the Province of Ontario through the Ministry of Research and Innovation. M.R.\ and J.B.\ were partly supported through a Projectruimte grant of the Netherlands Organisation for Scientific Research (NWO).

\appendix
\begin{appendices}
\section{The Bochner technique}
\label{app:bochner}

Eq. \eqref{eq:killing-energy-hodged} can be derived with a variation on what is known as the Bochner technique. First, it is necessary to make a choice for the coefficients in the definition of the inner product $|\cdot,\cdot|$ on $k$-forms. The coefficients are fixed by requiring that the exterior derivative $d$ and the co-differential $\delta$ are adjoint to each other with respect to $|\cdot,\cdot|$, i.e. $|A,\delta B|=|dA,B|$. Taking the correct coefficients into account, the contraction of the exterior derivative $d\omega$ with itself and the square of the co-differential $\delta \omega$ in covariant derivatives $\nabla_{\mu} \omega_\nu$ of a one-form $\omega$ are given by
\begin{align}
|d\omega,d\omega|&=\frac{1}{2}(d\omega)_{\mu \nu}(d\omega)^{\mu \nu} =2\nabla_{[\mu}\omega_{\nu]} \notag \\
|\delta \omega,\delta \omega|&= \delta \omega^2= \nabla_\mu\omega^\mu \nabla_\nu \omega^\nu. \notag
\end{align}
For arbitrary dimension $n$ we can expand the contraction of $K_{\mu \nu}=2\nabla_{(\mu}\omega_{\nu)}$ with itself in the covariant derivatives $\nabla_{\mu}\omega_\nu$,
\begin{align}
|K,K| =&\ 2(\nabla_{[\mu}\omega_{\nu]} \!+\! \nabla_\nu \omega_\mu)(\nabla^{[\mu}\omega^{\nu]} \!+\! \nabla^\nu \omega^\mu) \notag\\
=&\ 2(\nabla_{[\mu}\omega_{\nu]}\nabla^{[\mu}\omega^{\nu]} \!+\! \nabla_\nu \omega_\mu\nabla^\mu \omega^\nu) \notag\\
=&\ 2(\nabla_{[\mu}\omega_{\nu]}\nabla^{[\mu}\omega^{\nu]} \!-\!\omega_\mu\nabla_\nu \nabla^\mu \omega^\nu+\nabla_\nu( \omega_\mu\nabla^\mu \omega^\nu)) \notag\\
=&\ 2(\nabla_{[\mu}\omega_{\nu]}\nabla^{[\mu}\omega^{\nu]} \!-\! \omega^\mu\nabla_{[\nu} \nabla_{\mu]} \omega^\nu\! -\!  \omega^\mu\nabla_{\mu} \nabla_{\nu} \omega^\nu \!+\!\nabla_\nu( \omega_\mu\nabla^\mu \omega^\nu)) \notag\\
=&\ 2(\nabla_{[\mu}\omega_{\nu]}\nabla^{[\mu}\omega^{\nu]} \!-\! \omega_\mu R^\kappa_{\hphantom{\kappa}\nu\kappa\mu} \omega^\nu \!+\!  \nabla_\mu\omega^\mu \nabla^{\nu} \omega^\nu \notag\\
&-\!  \nabla_\mu(\omega^\mu \nabla_{\nu} \omega^\nu)\!+\!\nabla_\nu( \omega_\mu\nabla^\mu \omega^\nu)) \notag\\
=&\ |d\omega,d\omega| \!+\! 2 |\delta\omega,\delta\omega| \!-\! 2R_{\mu\nu} \omega^\mu \omega^\nu \! -\! 2 \nabla_\mu(F^\mu) \notag\\
\label{appKillingEquationNorm1}
\end{align}
where $F^\mu=\omega^\mu \nabla_{\nu} \omega^\nu-\omega_\nu\nabla^\nu \omega^\mu$. We use the inner product for scalars $|\phi,\phi|=\phi^2$, one-forms $|\omega,\omega|=\omega_{\mu}\omega^{\mu}$ and two-forms $|\psi,\psi|=\frac{1}{2}\psi_{\mu\nu}\psi^{\mu\nu}$. The last term is a total derivative of $F^\mu$ and will therefore not play a role in manifolds $\mathcal{M}$ without boundary. With these results we can rewrite the Killing energy $E(\omega)=2\int|K,K|$ with respect to a generalisation of the ordinary inner product $|\cdot,\cdot|$ on a closed manifold $\mathcal{M}$ in the following way,
\begin{equation}
E(\omega) = \int_\mathcal{M} dV \ \left( 2| d \omega,d\omega| + 4|\delta\omega,\delta\omega|-4 Ric(\omega,\omega) \right) .
\label{appKillingEnergyHodge}
\end{equation}
From now on we assume that the manifold $\mathcal{M}$ is two-dimensional such that the Ricci tensor $R_{\mu \nu}$ simplifies to $R_{\mu \nu}= \frac{R}{2} g_{\mu\nu}$, where $R$ is the Ricci scalar. The operators $d$ and $\delta$ are the adjoints of each other with respect to the inner product $|\cdot,\cdot |$. We can therefore write the Killing energy $E(\omega)$ very compactly,
\begin{equation}
E(\omega) = \int_\mathcal{M} dV \ \left( 2| \delta d \omega, \omega | +4 | d \delta \omega, \omega | - 2 R | \omega,\omega | \right).
\label{appKillinEnergyAdjoint}
\end{equation}
Or in a shorthand form introducing the operator $S$,
\begin{equation}
E = \int_\mathcal{M} dV \ | S \omega, \omega |,
\label{appKillinEnergyS}
\end{equation}
with
\begin{equation}
S = 2\delta d + 4 d \delta - 2R.
\label{appKillingOperator}
\end{equation}
\end{appendices}

\end{document}